\documentclass[aps,prx,reprint,superscriptaddress,amsmath,amssymb,floatfix]{revtex4-2}
\usepackage{graphicx}
\usepackage{bm}
\usepackage{booktabs}
\usepackage{array}
\usepackage{tabularx}
\usepackage{siunitx}
\usepackage{xcolor}
\usepackage[colorlinks=true,allcolors=blue]{hyperref}
\makeatletter
\renewcommand{\frontmatter@thefootnote}{%
  \ifnum\value{mpfootnote}=1\relax
    \ensuremath{\dagger}
  \else
    \@fnsymbol{\value{mpfootnote}}%
  \fi
}
\makeatother

\begin{document}
\title{Device Model Linking Singlet and Triplet Spin-State Properties to Organic Photovoltaic Device Characteristics}
\author{Jun Du}
\affiliation{Guangdong Basic Research Center of Excellence for Aggregate Science, School of Science and Engineering, The Chinese University of Hong Kong (Shenzhen), Shenzhen, Guangdong 518172, China}
\author{Liming Liu}
\affiliation{Guangdong Basic Research Center of Excellence for Aggregate Science, School of Science and Engineering, The Chinese University of Hong Kong (Shenzhen), Shenzhen, Guangdong 518172, China}
\author{Jun Yan}
\thanks{\texorpdfstring{Corresponding author. }{Corresponding author. }\href{mailto:yanjun@cuhk.edu.cn}{yanjun@cuhk.edu.cn}}
\affiliation{Guangdong Basic Research Center of Excellence for Aggregate Science, School of Science and Engineering, The Chinese University of Hong Kong (Shenzhen), Shenzhen, Guangdong 518172, China}
\begin{abstract}

    This work proposes an exciton-kinetics coupled drift--diffusion (EKDD)
    model that explicitly bridges device-level charge transport and spin
    singlet and triplet excited-state kinetics. State populations and
    reversible photophysical transitions are solved self-consistently,
    enabling direct calculations of both transient and steady-state device
    responses. We systematically investigate how local exciton state --
    charge-transfer state (LE-CT), charge-transfer state -- charge-separated
    state (CT--CS), and local exciton state -- localized triplet state
    (LE--T) energy offsets, alongside triplet-transfer kinetics, govern
    charge generation, recombination, voltage loss, and charge extraction in
    organic photovoltaics. We identify CT-CS offset as the primary factor
    controlling charge generation and voltage losses, which has been largely
    overlooked. A positive offset makes CT3--CS separation exothermic and
    improves charge generation, but lowers the CS energy and increases
    voltage loss, whereas a negative offset makes separation endothermic,
    making charge generation highly sensitive to temperature. The model
    reproduces triplet-mediated recombination in PM6:Y6 and
    material-dependent triplet recycling in D18:NFA blends. We propose design
    rules for balanced charge generation and voltage loss via two charge
    transfer channels: the LE--CT--CS pathway favors a moderate LE--CT offset
    of about 0.15--0.20 eV and a weakly exothermic CT--CS offset of about
    0.05--0.075 eV; the T--CT3--CS pathway requires the same CT--CS alignment
    together with a small CT--T offset, approximately no larger than 0.1 eV,
    corresponding to a recycling factor $K\gtrsim{}10^{-2}$, to
    maintain rapid T to CT3 recycling. These results identify CT--CS
    alignment as the main determinant of the charge-generation--voltage-loss
    trade-off and triplet-recycling kinetics as the condition that determines
    whether this balance survives at the device level. This work provides a
    unified theoretical description from exciton generation to free charge
  collection in a device level, offering a convenient tool for both transient
    and steady state analysis in organic photovoltaics.
  
\end{abstract}
\maketitle

  \section{Introduction}

  Organic photovoltaics (OPVs) convert sunlight into electrical current
  through a sequence of coupled photophysical and charge-transport processes.
  The low dielectric constant of organic semiconductors produces strongly
  bound electron--hole pairs, known as excitons; therefore efficient operation
  requires interconnected multi-step processes involving exciton generation,
  charge separation, carrier transport, and extraction rather than charge
  transport alone\cite{xu2008photovoltaic},\cite{vandewal2016interfacial}.
  In a donor--acceptor blend, photoexcitation first creates a local exciton
  (LE). The LE exciton diffuses to a donor--acceptor interface and forms an
  interfacial charge-transfer (CT) state, in which the electron and hole
  reside mainly on different molecular components. CT state excitons can
  dissociate into charge-separated (CS) states, return to an LE, or recombine
  to the ground state.

  CT state sits in between LE and CS state, therefore is at the heart of
  OPV operation. Its energy, electronic coupling, wave-function
  delocalization, and energetic disorder influence both charge generation and
  recombination \cite{vandewal2016interfacial}. A higher CT energy generally
  reduces nonradiative voltage loss through the energy-gap law
  \cite{benduhn2017intrinsic},\cite{azzouzi2018nonradiative}. However,
  reducing the LE--CT energy offset can also reduce charge-separation
  efficiency, making the LE lifetime and interfacial kinetics important
  \cite{classen2020role},\cite{azzouzi2022}. A quantitative description must
  therefore account for the competition between CT dissociation, geminate
  recombination, and nongeminate recombination.

  Spin-triplet states are another nonradiative loss channel because of their
  lower-lying energies.
  Nongeminate electron--hole recombination produces singlet and
  triplet CT states (CT1 and CT3) in an approximate 1:3 ratio, although the
  branching depends on molecular coupling and spin mixing
  \cite{xu2008photovoltaic}. The photophysics of triplets, including their
  formation, low energies, long lifetimes, and weak radiative emission, has
  been reviewed extensively \cite{kohler2009triplet}. In OPV blends, CT3 can
  dissociate into free carriers or undergo back charge transfer to a
  lower-energy localized triplet exciton (T). Because the T-to-ground-state
  transition is spin-forbidden, T commonly decays slowly radiatively and can act as a
  trap or a terminal nonradiative loss channel. Quantitative spectroscopy
  showed that bimolecular recombination can populate triplets and limit the
  internal quantum efficiency\cite{chow2014quantitative}; in PM6:Y6,
  recombination through non-emissive acceptor triplets was assigned to a
  large fraction of the open-circuit voltage loss\cite{gillett2021role}.

  Triplets may form by geminate or nongeminate recombination, or by
  intersystem crossing, depending on the materials and morphology, and are
  not necessarily irreversible losses\cite{rao2013spin}.
  The relative alignment of the T and CT3 levels does not alone determine the nonradiative
  voltage loss \cite{benduhn2018impact}, and morphology can suppress triplet formation by
  reducing nongeminate encounters\cite{jiang2022suppressed}. Triplet states
  can also assist charge generation when singlet--triplet splitting is small
  \cite{yamanaka2022significant}. Recent work indicates that a localized T
  can transfer back to CT3 before decaying, and that the resulting CT3
  population can redissociate into extractable carriers when the energy
  offset and transfer rates are favorable\cite{shadabroo2025triplet},\cite{li2026recycling}.
  Thus, the fate of triplet excitons is governed by the coupled rates of CT3
  formation, CT3 to T transfer, T redissociation, and triplet decay, together
  with carrier transport.

  Rate equations are widely adopted to quantitatively describe excited-state
  transition \cite{azzouzi2022}, \cite{li2026recycling},
  \cite{langentepe2026dark}. However, standalone rate models cannot directly
  reproduce device electrical responses and ignore the influence of electric
  fields. For this reason, coupling excited-state rate equations with
  free-carrier drift--diffusion transport has become a standard approach,
  enabling device-level simulations based on microscopic excited-state
  kinetics. Koster et al. introduced one of the first device-model examples
  of coupled excitonic and CS dynamics using Onsager-Braun theory, yet,
  focusing only on CT state decay \cite{koster2005device}. Kirchartz et al.
  later developed a more complete model that interconnects LE, CT, and CS
  states, allowing direct calculation of current--voltage responses from
  excited-state transitions \cite{kirchartz2008electro}. A similar method was
  adapted by Giebink et al., leading to an analytical description of
  current-voltage characteristics. Nevertheless, all previous models focused
  on steady-state responses, with no explicit investigations of transient
  dynamics of exciton and free charge \cite{giebink2010ideal}. Azzouzi et al.
  made an important step toward a unified description by reconciling
  interfacial-state kinetics with device performance and by quantifying how
  LE--CT and CT--CS energy offsets as well as their excited state properties
  affect transient responses and power-conversion
  efficiency\cite{azzouzi2022}. Despite these advances, a device model that
  interconnects LE, CT1, CT3, T, and CS, including reversible triplet
  recycling, with spatial electrostatics and carrier transport is still
  missing.

  Here, we develop an exciton-kinetics coupled drift--diffusion (EKDD) model
  that accounts for transitions between multiple state, including LE, CT1,
  CT3, T, and CS. Reversible kinetic transitions are solved together with the
  Poisson and carrier drift--diffusion equations, allowing direct
  calculations of transient and steady-state device responses. The framework
  provides a direct connection between excited-state properties, microscopic
  kinetics, and measurable current--voltage characteristics and photovoltaic
  metrics. We identify CT-CS transition as an important but previously
  overlooked factor for charge generation. The results show that exothermic
  CT3 to CS transfer increases dissociation but lowers the CS energy and
  increases voltage loss, whereas endothermic transfer preserves energy but
  makes charge generation dependent on CT3 dissociation and T-to-CT3
  recycling. We test our model via two case studies focusing on triplet
  terminal recombination in PM6:Y6 and triplet recycling in D18:NFA blends.
  Finally, we deliver practical design rules from two charge generation
  pathways, i.e. LE-CT-CS and T-CT3-CS, that balance charge generation,
  voltage preservation, and rapid triplet recovery.

  \section{Theory and Model}

  \subsection{Exciton kinetics Coupled with drift-diffusion model}

  Organic solar cells (OSC) operate through a series of microscopic processes
  as shown in Fig.1. Photon absorption generates strongly bound singlet
  excitons at the local singlet exciton (LE) level. These LE excitons diffuse
  to the donor--acceptor interface and undergo charge transfer to form
  singlet charge-transfer (CT1) excitons; in the meanwhile, LE excitons may
  also decay radiatively or nonradiatively to the ground state
  $S_{0}$. CT1 states can further dissociate into free
  charge-separated (CS) carriers, back-transfer to reform LE excitons, or
  decay directly to $S_{0}$. After free charges are generated, they
  either transport to the electrodes or undergo nongeminate bimolecular
  recombination to repopulate the CT in a 1:3 spin-statistical ratio: one
  quarter form singlet CT1 states and three quarters form triplet CT3 states.
  Formed CT1 excitons may redissociate, back-transfer to LE, or decay to
  the ground state. CT3 excitons can either redissociate into free carriers
  or relax into localized triplet (T) states. Direct CT3 decay to
  $S_{0}$ is spin-forbidden and neglected. Importantly, T excitons
  may redissociate back to CT3, followed by redissociation into free charges
  at moderate energy offsets between T and CT3, or decay nonradiatively to
  $S_{0}$ as the dominant loss channel. Finally, the transport of
  free charges across the active layer governs charge extraction and
  macroscopic current--voltage behavior.

  A complete model capturing the full process from exciton generation to free
  charge transport requires connections between exciton and free charge
  kinetics, which are often modeled separately, via either exciton
  kinetic equations\cite{langentepe2026dark} or drift and diffusion
  equations\cite{koster2005device}, therefore only address part of the
  physics. Kinetic rate equations capture excitonic state dynamics and loss
  mechanisms but cannot directly calculate real device characteristics such
  as current-voltage curves or fill factor\cite{langentepe2026dark}.
  Conventional drift--diffusion models describe macroscopic charge transport
  but treat recombination as an empirical parameter, without explicit
  connection to exciton, charge-transfer, and triplet
  dynamics\cite{koster2005device}. A model that couples exciton kinetics to
  free charge dynamics is rare and so far no models have included triplet
  states\cite{kirchartz2008electro},\cite{azzouzi2022}. We therefore develop
  here a self-consistent unified framework that couples singlet, triplet,
  charge transfer, charge transfer triplet, with free charge drift and
  diffusion transport, governed by the following set of equations:

  \begin{widetext}
\small
\begin{align}
\partial{}_{t}\left[LE\right] & = G_{LE}
    + k_{trans}^{CT1\mathbin{\rightarrow}LE}\left[CT1\right] -
    k_{trans}^{LE\mathbin{\rightarrow}CT1}\left[LE\right] -
    k_{rec}^{LE} \left(\left[LE\right]-\left[LE\right]_{0}\right) \\
\partial{}_{t}\left[CT1\right] & = k_{trans}^{LE\mathbin{\rightarrow}CT1}
    \left[LE\right]+B_{for}^{CS\mathbin{\rightarrow}CT}  \left(n
    p-n_{i}^{2}\right)-\left(k_{trans}^{CT1\mathbin{\rightarrow}LE}+k_{diss}^{CT1\mathbin{\rightarrow}CS}\right)\left[CT1\right]
    - k_{rec}^{CT1} \left(\left[CT1\right]
    -\left[CT1\right]_{0}\right) \\
\partial{}_{t}\left[CT3\right] & = k_{trans}^{T\mathbin{\rightarrow}CT3}\left[T\right]-\left(k_{diss}^{CT3\mathbin{\rightarrow}CS}+k_{trans}^{CT3\mathbin{\rightarrow}T}\right)\left[CT3\right]+3
    B_{for}^{CS\mathbin{\rightarrow}CT} \left(n
    p-n_{i}^{2}\right) \\
\partial{}_{t}\left[T\right] & = k_{trans}^{CT3\mathbin{\rightarrow}T}\left[CT3\right]
    - k_{trans}^{T\mathbin{\rightarrow}CT3}\left[T\right] -
    k_{rec}^{T}\left(\left[T\right]-\left[T\right]_{0}\right) \\
\partial{}_{t}n & = \nabla{}
    \cdot{}  \mathbf{J}_{n}+k_{diss}^{CT1\mathbin{\rightarrow}CS}\left[CT1\right]
    + k_{diss}^{CT3\mathbin{\rightarrow}CS}\left[CT3\right] - 4
    B_{for}^{CS\mathbin{\rightarrow}CT} \left(n
    p-n_{i}^{2}\right) \\
\partial{}_{t}
    p & = -\nabla{} \cdot{}
    \mathbf{J}_{p}+k_{diss}^{CT1\mathbin{\rightarrow}CS}\left[CT1\right] +
    k_{diss}^{CT3\mathbin{\rightarrow}CS}\left[CT3\right] - 4
    B_{for}^{CS\mathbin{\rightarrow}CT} \left(n
    p-n_{i}^{2}\right) \\
0 & = 
    \nabla{}\cdot{} \left(\varepsilon{} \nabla{} V\right) +
    q\left(p - n + N_{d}-N_{a}\right)
\end{align}
\normalsize
\end{widetext}

  where $V$ is the electrostatic potential; $n$ and $p$ are
  the free electron and hole densities, respectively; $\left[LE\right],
  \left[CT1\right], \left[CT3\right]$ and $\left[T\right]$ denote the
  densities of local singlet excitons, singlet charge-transfer excitons,
  triplet charge-transfer excitons, and localized triplet excitons,
  respectively; $\varepsilon{}$ is the permittivity; $q$ is the
  charge per electron; $N_{d}$ and $N_{a}$ denote the donor
  concentration and acceptor concentration; $\mathbf{J}_{n}= n
  \mu{}_{n}\nabla{} E_{Fn}$ and $\mathbf{J}_{p} = p
  \mu{}_{p} \nabla{} E_{Fp}$ are the current of electrons and holes
  expressed by quasi-fermi level $E_{Fn}$ and $E_{Fp}$,
  respectively, $\mu{}_{n}, \mu{}_{p}$ are electron and hole
  mobility; $n_{i}$ is intrinsic density; $G_{LE}$ is the
  generation rate of LE state; $B_{for}^{CS\mathbin{\rightarrow}CT}$
  is the constant rate of formation of a CT1 or CT3 state in a 1:3
  spin-statistical ratio from free charge carriers, according to quantum
  mechanics; $k_{diss}^{CT3\mathbin{\rightarrow}CS}$ and
  $k_{diss}^{CT1\mathbin{\rightarrow}CS}$ are the dissociation rate
  constants of CT states; $k_{trans}^{LE\mathbin{\rightarrow}CT1}$ and
  $k_{trans}^{CT1\mathbin{\rightarrow}LE}$ are the transfer rate
  constant between CT1 and LE states; $k_{trans}^{T\mathbin{\rightarrow}CT3}$
  and $k_{trans}^{CT3\mathbin{\rightarrow}T}$ are the  transfer rate
  constants between CT3 and T states; $k_{rec}^{LE} ,
  k_{rec}^{CT1}$ and $k_{rec}^{T}$ are the decay rate
  constants for LE, CT1 and T states recombination;
  $\left[LE\right]_{0}, \left[CT1\right]_{0}$ and
  $\left[T\right]_{0}$ are the equilibrium densities of the
  respective excitonic states.

\begin{figure*}[t]
\centering
\includegraphics[width=0.8\textwidth]{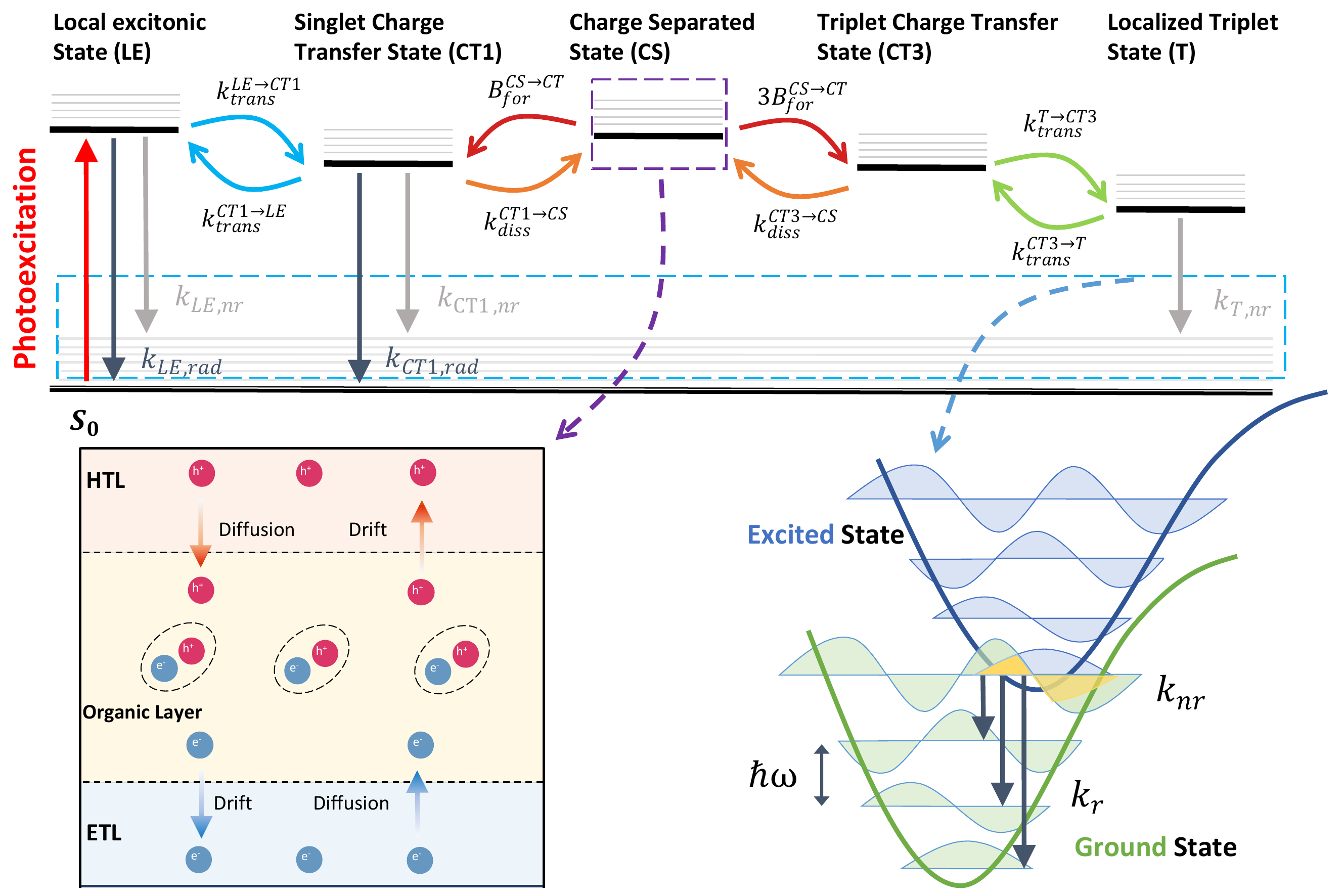}
\caption{\textbf{Schematic representation of the coupled excited-state and device processes included in the EKDD model.} Photon absorption promotes the ground state $S_{0}$ to a local exciton (LE), followed by reversible LE--CT1 interconversion. CT1 can dissociate into charge-separated (CS) state, while nongeminate CS recombination forms CT1 and CT3 with approximate 1:3 singlet--triplet branching. CT3 may dissociate into CS or transfer reversibly to a localized triplet exciton (T), which can either recycle through T to CT3 transfer or decay nonradiatively. Radiative and nonradiative relaxation pathways to $S_{0}$, carrier transport through the organic layer and extraction by the HTL and ETL, and the corresponding excited- and ground-state vibronic modes are illustrated schematically. Arrows indicate the allowed transitions, and $\hbar{} \omega{}$ denotes the dominant phonon energy.}
\label{schematic}
\end{figure*}

  We apply the following boundary conditions for voltage and current
  densities:

  \begin{align}
\left.V\right|_{x=0} & = 0, \\
\left.V\right|_{x=
    L} & = V_{bi} - V_{app} \\
\left.\mathbf{J}_{n}\right|_{x=0} & = s_{n,l}
    \left(\left.n\right|_{x=0}-n_{0,l}\right) \\
\left.\mathbf{J}_{n}\right|_{x=L} & = -s_{n,r}
    \left(\left.n\right|_{x=L}-n_{0,r}\right) \\
\left.\mathbf{J}_{p}\right|_{x=0} & = -s_{p,l}
    \left(\left.p\right|_{x=0}-p_{0,l}\right) \\
\left.\mathbf{J}_{p}\right|_{x=L} & = s_{p,r}
    \left(\left.p\right|_{x=L}-p_{0,r}\right)
\end{align}

  where $V_{bi}$ is built-in potential, $V_{app}$ is
  applied voltage, $s_{n/p, l/r}$ denote electrons/holes surface
  recombination velocity on left/right boundary, $n_{0,l/r}$ and
  $p_{0, l/r}$ are equilibrium electron and hole density on
  left/right boundary. For ordinary differential equations, their equilibrium
  densities are taken as the initial conditions. The model is schematically
  illustrated in Fig.\ref{schematic}. All equations are solved self-consistently
  using a recently developed numerical method\cite{du2026stable}.

  \subsection{Detailed balance for elementary transitions}

  We assume detailed balance within elementary processes, therefore forward
  and backward rates as well as the equilibrium densities of different
  species are interconnected for each transition\cite{azzouzi2022}. First, we
  define the energies of the five states: $E_{LE}. E_{CT1},
  E_{CT3}, E_{T}$ and $E_{CS}$. Here $E_{CS}$
  equals the transport band gap $E_{g}$. The singlet and triplet
  charge-transfer states (CT1 and CT3) are assumed to be
  energetically degenerate, $E_{CT1} = E_{CT3} = E_{CT}$,
  because spatially separated CT excitons have very small exchange splitting.
  The relevant energy offsets are defined as $\Delta{}E_{LE-CT} =
  E_{LE} - E_{CT}$ and $\Delta{}E_{CT-T} =E_{CT} -
  E_{T}$, with the singlet--triplet splitting of the acceptor given by
  $\Delta{}E_{ST} = E_{LE} - E_{T} = \Delta{}E_{LE-CT}
  + \Delta{}E_{CT-T}$. Each localized state has an effective density of
  states: $N_{LE}, N_{CT1}$ and $N_{T}$. Triplet CT has
  three times as many states as the singlet CT, so $N_{CT3} =
  3N_{CT1}$. Since localized triplet excitons only form at
  donor--acceptor interfaces from $CT3$ relaxation, they occupy the same
  sites as $CT3$, giving $N_{CT3} = N_{T}$
  \cite{langentepe2026dark}. At thermal equilibrium, the population of each
  state follows the Boltzmann distribution:

  \begin{align}
\left[X\right]_{0} = N_{X} \exp \left(-\frac{E_{X}}{k_{B}T}\right),
\end{align}

  where $\left[X\right]_{0}$ is the equilibrium density of state
  $X$, $N_{X}$ and $E_{X}$ are corresponding effective
  density of states (eDOS) and energy, $k_{B}$ is Boltzmann constant,
  and $T$ is the temperature.

  By the principle of detailed balance, the forward and backward rates
  between any two states are linked by their equilibrium populations. For
  states A and B,

  \begin{align}
\frac{k_{A\mathbin{\rightarrow}B}}{k_{B\mathbin{\rightarrow}A}} =
    \frac{\left[B\right]_{0}}{\left[A\right]_{0}}.
\end{align}

  CT dissociation into free carriers can occur on a picosecond timescale in
  favorable blends, but its effective rate and yield need not be independent
  of the CT--CS energy offset\cite{azzouzi2022,muller2023dissociation}. A negative
  (endothermic) CT--CS offset can
  introduce a thermally activated barrier and allow CT recombination or
  relaxation to compete, whereas a positive offset favors dissociation but
  lowers $E_{CS}$ and can increase voltage loss\cite{azzouzi2022,vezie2019,
  muller2023dissociation,classen2020role}. In the present reference
  calculations, we use a fixed elementary dissociation rate while the
  offset-dependent reverse and formation rates are constrained by detailed
  balance. Following experimental observations \cite{gillett2021role} and due
  to $E_{CT1}\cong{} E_{CT3}$, we set both CT dissociation rates to the same constant value
  $k_{diss}^{CT1\mathbin{\rightarrow}CS} =
  k_{diss}^{CT3\mathbin{\rightarrow}CS} = 10^{11} s^{-1}$. Using
  detailed balance between CT and CS states, we find the bimolecular CT
  formation rate $B_{for}^{CS\mathbin{\rightarrow}CT}=k_{diss}^{CT1\mathbin{\rightarrow}CS}
  \frac{\left[CT1\right]_{0}}{n_{i}^{2}}$ \cite{azzouzi2022},
  where $n_{i}$ is intrinsic free-carrier density. The 1:3
  spin-statistical ratio between CT1 and CT3 still holds under detailed
  balance, because the factor of three in $N_{CT3}$ is exactly
  reflected in the equilibrium populations.

  Forward transition rates from LE to CT1 and from CT3 to T are calculated
  from Marcus theory \cite{marcus1985electron}:

  \begin{widetext}
\small
\begin{align}
k\left(\Delta{}E\right) = \frac{H_{a
    b}^{2}}{\hbar{}}\sqrt{\frac{\pi{}}{\lambda{} k_{B}
    T}}\exp \left[-\frac{\left(\lambda{}+\Delta{}E\right)^{2}}{4
    \lambda{} k_{B}T}\right],
\end{align}
\normalsize
\end{widetext}

  where $H_{a b}$ is the electronic coupling, $\lambda{}$ is
  the reorganization energy, and $\Delta{}E$ is the driving force. We
  compute the forward rates $k_{trans}^{LE\mathbin{\rightarrow}CT1}$
  and $k_{trans}^{CT3\mathbin{\rightarrow}T}$ by inserting the driving
  forces $-\Delta{}E_{LE-CT}$ and $-\Delta{}E_{CT-T}$.
  The reverse rates are then obtained directly from detailed balance:
  \begin{equation*}
  \begin{aligned}
  k_{trans}^{CT1\mathbin{\rightarrow}LE}/k_{trans}^{LE\mathbin{\rightarrow}CT1}
    &= \left[LE\right]_{0}/\left[CT1\right]_{0},\\
  k_{trans}^{T\mathbin{\rightarrow}CT3}/k_{trans}^{CT3\mathbin{\rightarrow}T}
    &= \left[CT3\right]_{0}/\left[T\right]_{0}.
  \end{aligned}
  \end{equation*}
  This approach reduces the number of independent parameters: only the
  forward rates require Marcus calculations, and all reverse rates follow
  automatically from thermodynamic consistency. A Summary sheet can be found
  at Table 1.
\begin{table*}[t]
\caption{Sequence of calculation for input parameters.}
\centering
\footnotesize
\setlength{\tabcolsep}{4pt}
\renewcommand{\arraystretch}{1.12}
\resizebox{\textwidth}{!}{
\begin{tabular}{lll}
\toprule
Input & Expression & Output \\
\midrule
$N_{c},
    N_{v}, E_{g}$ & $n_{i}^{2} = N_{c}
    N_{v} \exp \left(-E_{g}/k_{B}T\right)$ & $n_{i}^{2}$ \\
$N_{CT1}$ & $N_{CT3}
    = 3N_{CT1}$ & $N_{CT3}$ \\
$N_{CT3}$ & $N_{T}
    = N_{CT3}$ & $N_{T}$ \\
$E_{CT}$ & $\left[CT1\right]_{0}=N_{CT1}
    \exp \left(-E_{CT}/k_{B}T\right)
    $ & $\left[CT1\right]_{0}$ \\
$\left[CT1\right]_{0}$ & $\left[CT3\right]_{0}
    = 3\left[CT1\right]_{0}$ & $\left[CT3\right]_{0}$ \\
$E_{g},
    E_{CT}$ & $\Delta{}E_{CT-CS} =
    E_{CT}-E_{g}$ & $\Delta{}E_{CT-CS}$ \\
$n_{i}^{2},
    k_{diss}^{CT1\mathbin{\rightarrow}CS}$ & $\left[CT1\right]_{0}/n_{i}^{2}=B_{for}^{CS\mathbin{\rightarrow}CT}/k_{diss}^{CT1\mathbin{\rightarrow}CS}$ & $B_{for}^{CS\mathbin{\rightarrow}CT}$ \\
$k_{diss}^{CT1\mathbin{\rightarrow}CS}$ & $k_{diss}^{CT3\mathbin{\rightarrow}CS}=k_{diss}^{CT1\mathbin{\rightarrow}CS}$
     & $k_{diss}^{CT3\mathbin{\rightarrow}CS}$ \\
$E_{CT},E_{LE}
    $ & $\Delta{}E_{LE-CT} =
    E_{LE}-E_{CT}$ & $\Delta{}E_{LE-CT}$ \\
$E_{LE},
    E_{T} $ & $\Delta{}E_{LE-T}
    = E_{LE} -E_{T}$ & $\Delta{}E_{LE-T}$ \\
$E_{CT},
    E_{T}$ & $\Delta{}E_{CT-T}=E_{CT}-E_{T}$ & $\Delta{}E_{CT-T}$ \\
$\Delta{}E_{LE-CT},
    N_{LE}, N_{CT1},
    \left[CT1\right]_{0}$ & $\left[LE\right]_{0}/\left[CT1\right]_{0}=(N_{LE}/N_{CT1})\exp \left[-\Delta{}E_{LE-CT}/(k_{b}T)\right]$ & $\left[LE\right]_{0}$ \\
$\Delta{}E_{CT-T},N_{CT3},
    N_{T}, \left[CT3\right]_{0}$ & $\left[T\right]_{0}/\left[CT3\right]_{0}=(N_{T}/N_{CT3})\exp \left[\Delta{}E_{CT-T}/(k_{b}T)\right]$ & $\left[T\right]_{0}$ \\
$-\Delta{}E_{LE-CT},
    -\Delta{}E_{CT-T}$ & $k\left(\Delta{}E\right)=\frac{H_{ab}^{2}}{\hbar{}}\sqrt{\frac{\pi{}}{\lambda{}
    k_{B}T}}\exp \left\{-\frac{\left(\lambda{}+\Delta{}E\right)^{2}}{4
    \lambda{} k_{B}T}\right\}$ & $k_{trans}^{LE\mathbin{\rightarrow}CT1},
    k_{trans}^{CT3\mathbin{\rightarrow}T}$ \\
$\left[LE\right]_{0},\left[CT1\right]_{0},
    k_{trans}^{LE\mathbin{\rightarrow}CT1}$ & $k_{trans}^{CT1\mathbin{\rightarrow}LE}/k_{trans}^{LE\mathbin{\rightarrow}CT1}
    =\left[LE\right]_{0}/\left[CT1\right]_{0}$ & $k_{trans}^{CT1\mathbin{\rightarrow}LE}$ \\
$\left[T\right]_{0},\left[CT3\right]_{0},k_{trans}^{CT3\mathbin{\rightarrow}T}$ & $k_{trans}^{T\mathbin{\rightarrow}CT3}/k_{trans}^{CT3\mathbin{\rightarrow}T}=\left[CT3\right]_{0}/\left[T\right]_{0}$ & $k_{trans}^{T\mathbin{\rightarrow}CT3}$ \\
\bottomrule
\end{tabular}}
\vspace{2pt}
\parbox{\textwidth}{\footnotesize \textsuperscript{*}\,
    Coupling strength $H_{a b}=0.01\,\mathrm{eV}$, Reorganization energy
    $\lambda{}\approx{}0.5\,\mathrm{eV}$, Planck constant $\hbar
    \approx{} 6.582\times{}10^{-16}\,\mathrm{eV}\,\mathrm{s}$ }
\end{table*}

  \subsection{Theory for exciton decay}

  At the donor-acceptor heterojunction, localized singlet excitons (LE),
  singlet charge-transfer states (CT1), and localized triplet excitons (T)
  can all decay to the electronic ground state $S_{0}$ via radiative
  and nonradiative channels. The total recombination rate constant for each
  excited state $Xin\left\{LE, CT1, T\right\}$ is given by

  \begin{align}
k_{rec}^{X} = k_{r}^{X} + k_{n r}^{X},\label{rec}
\end{align}

  where $k_{r}^{X}$ and $k_{n r}^{X}$ denote the
  radiative and nonradiative decay rate constants, respectively.
  Specifically, nonradiative decay proceeds via isoenergetic transitions
  between vibrational level of the initial and final electronic states,
  followed by fast and irreversible dissipation of excess vibrational energy
  to the surroundings\cite{kohler2026electronic}.

  Within the electric-dipole approximation, the energy-resolved spontaneous
  emission rate follows from Fermi's golden rule
  \cite{azzouzi2018nonradiative}:

  \begin{align}
k_{r}^{X}\left(E\right) = \frac{1}{3 \pi{}
    \varepsilon{}_{0} \hbar{}^{4}}\left(\frac{E}{c}\right)^{3}
    \left(\mu_{i f}^{X}\right)^{2}
    \mathrm{FCWD}^{X}\left(E\right).
\end{align}

  where $E$ is the emitted photon energy, $\mu{}_{i
  f}^{X}$ is the transition dipole moment between the excited state and
  the ground state, and $\mathrm{FCWD}^{X}\left(E\right)$ is the
  Franck--Condon weighted density of states for emission, which encodes the
  vibrational structure of the emission spectrum. In general, the FCWD
  uniformly accounts for high-frequency intramolecular vibrational modes,
  treated quantum mechanically, and low-frequency environmental or solvent
  modes, treated classically as a Gaussian broadening. Adopting the Jortner
  form of the Marcus--Levich--Jortner model, the emission FCWD reads
  \cite{azzouzi2018nonradiative},\cite{englman1970energy}:

    \begin{widetext}
\small
\begin{align}
\mathrm{FCWD}^{X}\left(E\right) = \frac{1}{\sqrt{4
      \pi{} \lambda{}_{l} k_{B} T}}\sum _{m, n
      =0}^{\infty{}}\frac{e^{-S}S^{m-n}
      n!}{m!}\left[L_{n}^{\left(m-n\right)}\left(s\right)\right]^{2}\exp \left(-\frac{\left[E-\Delta{}G_{0}+\lambda{}_{l}+\left(m-n\right)\hbar{}
      \Omega\right]^{2}}{4 \pi{} \lambda{}_{l} k_{B} T
      }\right)\,\exp \left(-\frac{n\hbar{}\Omega}{k_{B}T}\right),
\end{align}
\normalsize
\end{widetext}

  The Huang--Rhys factor $S =\lambda{}_{v}/\hbar{} \Omega$
  quantifies the electron--vibration coupling strength, with
  $\lambda{}_{v}$ the inner (high-frequency) reorganization energy
  and $\hbar{}\Omega$ the vibrational quantum energy, typically
  around 0.15 eV for C--C stretching modes in organic semiconductors. The
  indices $m$ and $n$ label the vibrational quantum numbers of the
  final and initial states, respectively;
  $L_{n}^{\left(m-n\right)}\left(s\right)$ denotes the
  generalized Laguerre polynomial describing the vibrational wavefunction
  overlap. The last exponential factor accounts for the thermal Boltzmann
  population of initial vibrational levels. The outer (low-frequency)
  reorganization energy $\lambda{}_{l}$ enters through the Gaussian
  energy broadening kernel. The transition dipole moment is related to the
  experimentally accessible oscillator strength $f_{osc}^{X}$
  via

  \begin{align}
\left(\mu_{i f}^{X}\right)^{2} = \frac{3}{2}
    \frac{\hbar{}^{2}}{E_{avg} m_{e}} f_{osc}^{X},
\end{align}

  where $E_{avg}$ is the average emission energy and
  $m_{e}$ is the free-electron mass. LE states typically exhibit
  oscillator strengths of order unity, yielding fast radiative decay on the
  nanosecond timescale. CT states, by contrast, are characterized by strongly
  reduced oscillator strengths \cite{azzouzi2018nonradiative}
  ($f_{osc}^{X} \sim  10^{-4} -10^{-2}$) due to their
  spatially separated electron and hole character, resulting in much weaker
  radiative transitions. For triplet states, the
  $T\mathbin{\rightarrow}S_{0}$ transition is spin-forbidden, so that
  $k_{r}^{T}$ is negligible and triplet decay proceeds almost
  exclusively through nonradiative channels. Then, the total radiative rate
  constant is obtained by integration over the emission interval:

  \begin{align}
k_{r}^{X} = \int k_{r}^{X}\left(E\right) d E.
\end{align}

  Nonradiative recombination proceeds via multiphonon emission, whereby the
  electronic excitation energy is dissipated as vibrational heat. Following
  Marcus electron-transfer theory and Fermi's golden rule, the nonradiative
  rate constant reads \cite{azzouzi2018nonradiative},\cite{jortner1976temperature}:

  \begin{align}
k_{n r}^{X} = \frac{2 \pi{}}{\hbar{}}V_{X}^{2}
    \mathrm{FCWD}^{X}\left(0\right),
\end{align}

  where $V_{X}$ is the electronic coupling matrix element between
  the excited state and the ground state, and
  $\mathrm{FCWD}^{X}\left(0\right)$ is the Franck--Condon weighted density
  of states evaluated at zero energy mismatch, i.e., the density of final
  vibrational states that are resonant with the initial electronic energy. 

  The electronic coupling $V_{X}$ is obtained from the generalized
  Mulliken--Hush (GMH) expression, which relates it to the transition dipole
  moment and the static dipole-moment difference between the two electronic
  states \cite{azzouzi2018nonradiative},\cite{Hush1967IVT}:

  \begin{align}
V_{X} = \frac{\mu{}_{i f}^{X}
    E_{X}}{\sqrt{\Delta{}
    \boldsymbol{\mu{}}_{X}^{2}+4\left(\mu_{i f}^{X}\right)^{2}}},
\end{align}

  here, $E_{X}$ denotes the excitation energy of state $X$ and
  $\Delta{} \boldsymbol{\mu{}}_{X}$ is the
  difference in static dipole moment between the excited and ground states.
  CT states exhibit particularly large dipole-moment differences
  \cite{Hush1967IVT} (10--30 D) owing to their charge-separated nature.

  \section{Theoretical results}

  \subsection{Impact of energetic offsets}

  We systematically investigate the effect of three key energy offsets on
  device performance using EKDD. The reference energies were $E_{LE}
  = 1.5 eV$, $E_{CT} = 1.3 eV$ and $E_{T} = 1.15 eV$, while
  the drift--diffusion and transport parameters were fixed to those listed in
  Table 2. For each energy-offset, the forward transfer rates were
  recalculated using Marcus theory, the reverse rates were obtained from
  detailed balance, and the state-specific recombination rates were
  recalculated using Eq.\ref{rec}. The bimolecular CT-formation
  coefficient was recalculated using the fixed CT dissociation rate
  $k_{diss}^{CT\mathbin{\rightarrow}CS} = 10^{11} s^{-1}$. The
  resulting kinetic equations were coupled self-consistently to the
  one-dimensional drift--diffusion equations and solved together to generate
  J-V curve. The exciton generation rate under illumination was
  $G_{LE} =1.65 \times{}10^{22} cm^{-3}/s$, corresponding
  to one-sun excitation.

  In addition to the energy-offset scans, we performed sensitivity scans of
  the properties of each LE, CT and T state, including high-frequency
  reorganization energy $\lambda{}_{v}$, low-frequency
  reorganization energy $\lambda{}_{l}$, and static energetic
  disorder which was introduced through a Gaussian distribution with
  $\sigma{}=0.01-0.1 eV$. The reorganization energies were varied from
  0.05 to 0.30 eV. The following analysis focuses on the T-state parameter
  scans, with sensitivity of LE and CT state properties discussed in Appendix
  A and B. We now define voltage losses and internal quantum efficiency
  as follows

  \begin{align}
V_{loss} = E_{LE}/q - V_{oc},\,IQE = J_{sc} /
    J_{ideal},\label{iqe}
\end{align}

  where $J_{ideal} = q G_{LE} L$, for the 100 nm active layer in
  Table 2, $J_{ideal} = 26.44 mA/cm^{2}$. Thus, the IQE directly
  measures the fraction of the maximum photogenerated current that is
  collected at short-circuit condition.
\begin{table*}[t]
\caption{Parameters used for calculating the impact of energetic offsets on device responses.}
\centering
\scriptsize
\setlength{\tabcolsep}{4pt}
\renewcommand{\arraystretch}{1.12}
\resizebox{\textwidth}{!}{
\begin{tabular}{llllll}
\toprule

      \emph{Quantity}
     & 
      \emph{Symbol}
     & 
      \emph{ETL}
     & 
      \emph{PM6:Y6}
     & 
      \emph{HTL}
     & 
      \emph{Unit}
     \\
\midrule

      Layer thickness
     & 
      $d$
     & 
      10
     & 
      100
     & 
      10
     & 
      $\mathrm{nm}$
     \\

      Relative permittivity
     & 
      $\varepsilon{}_{r}$
     & 
      3
     & 
      3
     & 
      3
     & 
      $-$
     \\

      Electron mobility
     & 
      $\mu{}_{n}$
     & 
      $1\times{}10^{-2}$
     & 
      $2\times{}10^{-3}$
     & 
      $1\times{}10^{-4}$
     & 
      $\mathrm{cm}^{2}\,\mathrm{V}^{-1}\,\mathrm{s}^{-1}$
     \\

      Hole mobility
     & 
      $\mu{}_{p}$
     & 
      $1\times{}10^{-4}$
     & 
      $2\times{}10^{-3}$
     & 
      $1\times{}10^{-2}$
     & 
      $\mathrm{cm}^{2}\,\mathrm{V}^{-1}\,\mathrm{s}^{-1}$
     \\

      Conduction-band eDOS
     & 
      $N_{c}$
     & 
      $1\times{}10^{20}$
     & 
      $1\times{}10^{19}$
     & 
      $1\times{}10^{20}$
     & 
      $\mathrm{cm}^{-3}$
     \\

      Valence-band eDOS
     & 
      $N_{v}$
     & 
      $1\times{}10^{20}$
     & 
      $1\times{}10^{19}$
     & 
      $1\times{}10^{20}$
     & 
      $\mathrm{cm}^{-3}$
     \\

      Electron affinity
     & 
      $\Phi{}_{EA}$
     & 
      $-4.0$
     & 
      $-3.9$
     & 
      $-3.2$
     & 
      $\mathrm{eV}$
     \\

      Ionisation potential
     & 
      $\Phi{}_{IP}$
     & 
      $-6.0$
     & 
      $-5.3$
     & 
      $-5.2$
     & 
      $\mathrm{eV}$
     \\

      Equilibrium Fermi level
     & 
      $E_{F0}$
     & 
      $-4.32$
     & 
      $-$
     & 
      $-4.8$
     & 
      $\mathrm{eV}$
     \\

      Donor concentration
     & 
      $N_{A}$
     & 
      $1\times{}10^{18}$
     & 
      $-$
     & 
      $-$
     & 
      $\mathrm{cm}^{-3}$
     \\

      Acceptor concentration
     & 
      $N_{D}$
     & 
      $-$
     & 
      $-$
     & 
      $1\times{}10^{18}$
     & 
      $\mathrm{cm}^{-3}$
     \\

      Left boundary electron recomb. velocity
     & 
      $s_{n, l}$
     & 
      $1.0\times{}10^{5}$
     & 
      $-$
     & 
      $-$
     & 
      $\mathrm{m}^{2}\,\mathrm{s}^{-1}$
     \\

      Left boundary hole recomb. velocity
     & 
      $s_{p, l}$
     & 
      $1.0\times{}10^{5}$
     & 
      $-$
     & 
      $-$
     & 
      $\mathrm{m}^{2}\,\mathrm{s}^{-1}$
     \\

      Right boundary electron recomb. velocity
     & 
      $s_{n, r}$
     & 
      $1.0\times{}10^{7}$
     & 
      $-$
     & 
      $1.0\times{}10^{7}$
     & 
      $\mathrm{cm}^{2}\,\mathrm{s}^{-1}$
     \\

      Right boundary hole recomb. velocity
     & 
      $s_{p, r}$
     & 
      $1.0\times{}10^{7}$
     & 
      $-$
     & 
      $1.0\times{}10^{7}$
     & 
      $\mathrm{cm}^{2}\,\mathrm{s}^{-1}$
     \\
\bottomrule
\end{tabular}
}
\end{table*}

  \begin{figure*}[t]
\centering
\includegraphics[width=0.93\textwidth]{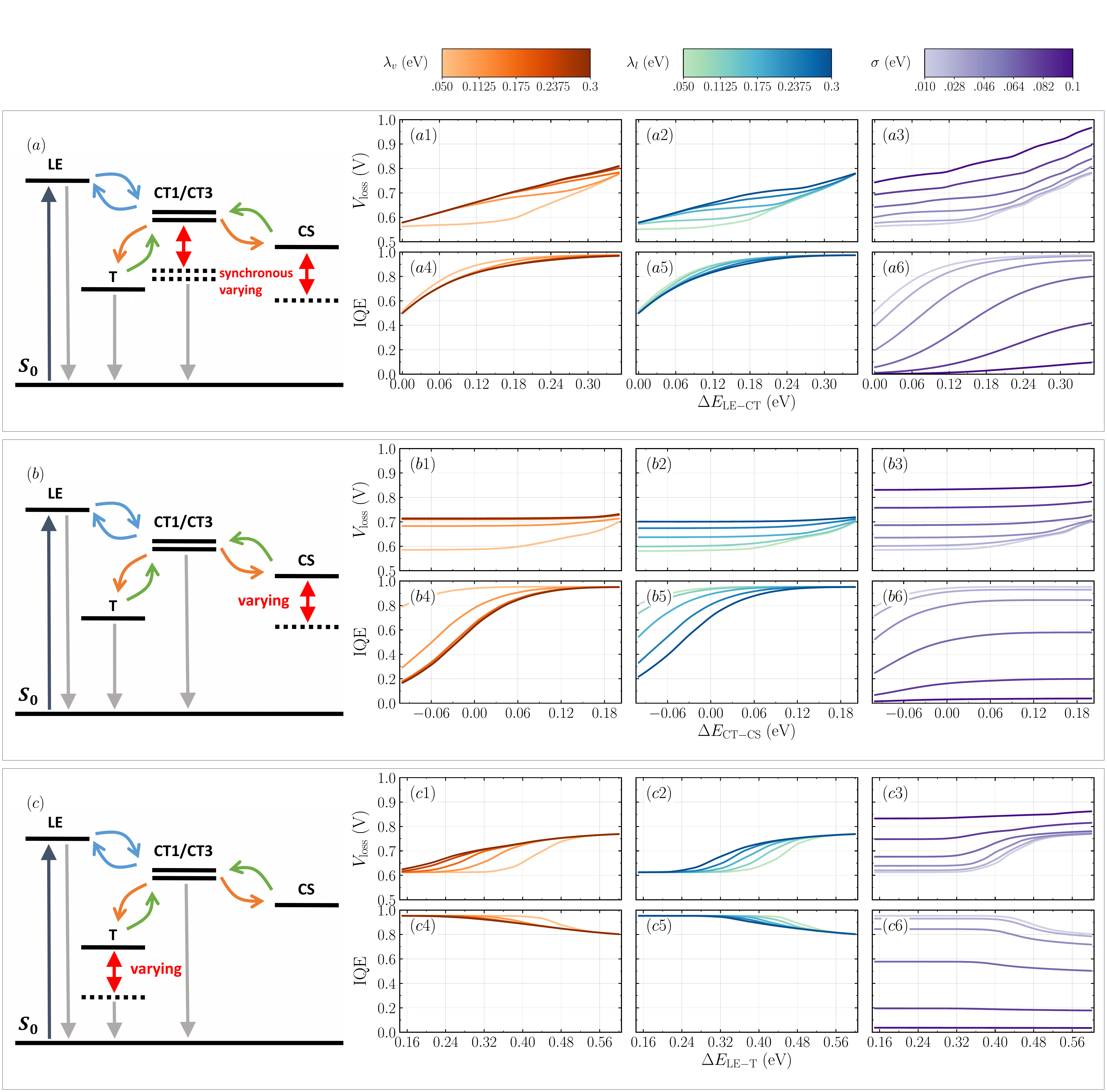}
\caption{Dependence of $V_{loss}$ and IQE on the three energy offsets and T-state properties. (a1)-(a6): LE--CT offset, (b1)-(b6): CT--CS offset, and (c1)-(c6): LE--T offset. The columns show variations in the T-state high-frequency reorganization energy $\lambda{}_{v}$, low-frequency reorganization energy $\lambda{}_{l}$, and static energetic disorder $\sigma{}$, respectively.}
\label{offset}
\end{figure*}

  \subsubsection{LE-CT offset}

  For the LE--CT scan, as shown in Fig.\ref{offset}(a1)-(a6),
  $E_{LE}$ was fixed at 1.50 eV, $E_{T}$ was fixed at 1.15
  eV, and the CT and CS energies were moved together according to
  $E_{CT} = E_{LE} - \Delta{}E_{LE-CT}$, thus the CT--CS
  offset was maintained at its reference value of 0.10 eV, while the T energy
  remained unchanged throughout the calculation.

  At small $\Delta{}E_{LE-CT}$, LE to CT transfer is slow and
  competes with LE recombination. Only a fraction of the photogenerated LE
  population reaches the CT, resulting in a low IQE. Increasing
  $\Delta{}E_{LE-CT}$ increases the LE to CT transfer rate and
  rapidly improves charge generation. The IQE therefore rises from
  approximately 0.5 at zero offset to approximately 0.95--0.98 at
  sufficiently large offset under low-disorder conditions, as shown in
  Fig.\ref{offset}(a4)-(a6).

  The improvement in IQE is accompanied by an increase in
  $V_{loss}$, see Fig.\ref{offset}(a1)-(a3). Because
  $E_{CT}$ and $E_{CS}$ are lowered together in the
  calculations done here, the energy available for sustaining the
  quasi-Fermi-level splitting decreases. Consequently, $V_{loss}$
  increases from approximately 0.56 V at small offsets to approximately
  0.78--0.81 V at large offsets for the low-disorder cases. A larger LE--CT
  driving force therefore improves charge generation but reduces the
  attainable voltage.

  The impact of reorganization energies is more pronounced in
  $V_{loss}$ than in IQE after efficient LE to CT transfer has been
  established. Variations in $\lambda{}_{v}$,
   $\lambda{}_{l}$, and $\sigma{}$ change the T-state decay
  and transfer kinetics, which modify the steady-state carrier population
  under open-circuit conditions. This directly changes the quasi-Fermi-level
  splitting and hence $V_{oc}$. Under short-circuit conditions,
  however, free carriers are rapidly extracted. As long as LE-to-CT
  conversion is efficient, moderate changes in the triplet properties have a
  smaller effect on IQE, hence the collected current. This different
  sensitivity indicates that triplet-state properties can contribute
  substantially to voltage loss without producing an equally large change in
  charge generation. Static disorder represents an important exception
  because sufficiently large $\sigma{}$ can also strongly suppress IQE
  by creating low-energy sub-states that act as kinetic traps and
  recombination centers.

  \subsubsection{CT-CS offset}

  For the CT--CS scan shown in Fig.\ref{offset}(b1)-(b6),
  $E_{LE}$, $E_{CT}$, and $E_{T}$ were fixed,
  while $E_{CS}$ was varied according to $E_{CS} =
  E_{CT} - \Delta{}E_{CT-CS}$. The LE and T energies therefore
  remained fixed at 1.50 and 1.15 eV, respectively. This scan isolates the
  competition between CT dissociation into CS and recombination from the CT.

  At small or negative $\Delta{}E_{CT-CS}$, CT dissociation is
  energetically weak or unfavorable and the charge separation process is
  endothermic. The CT population can then undergo recombination or reverse
  transfer before producing separated carriers. As a result, IQE is low and
  highly sensitive to the excited-state properties, as shown in
  Fig.\ref{offset}(b4)-(b6). Increasing $\Delta{}E_{CT-CS}$
  favors CT dissociation, and IQE approaches approximately 0.9--0.95 under
  favorable low-disorder conditions.

  The larger CT--CS driving force also reduces the attainable voltage.
  $V_{loss}$ increases from approximately 0.59 V at small offsets to
  approximately 0.70--0.73 V at large positive offset, see
  Fig.\ref{offset}(b1)-(b3). The CT--CS offset therefore introduces a
  second trade-off: a positive offset promotes free-carrier formation, but
  the associated lowering of $E_{CS}$ increases the voltage loss.

  The influence of the T-state reorganization energies on IQE is concentrated
  mainly in the small or negative CT--CS region. In this regime, CT
  dissociation is slow enough that CT recombination, triplet formation, and
  triplet recycling can compete with free-carrier generation. Changes in
  $\lambda{}_{v}$ and $\lambda{}_{l}$ then alter the
  relative populations of these pathways. At large positive offset, CT
  dissociation dominates and the IQE curves converge. Once the CT population
  rapidly reaches CS, changes in the T-state transfer rates have little
  opportunity to affect the collected current.

  The voltage loss remains more sensitive to T-state properties compared to
  IQE, even after the IQE approaches saturation. This is because
  recombination at open circuit condition continues to sample the CT and T,
  whereas charge collection becomes dominated by rapid CT dissociation and
  carrier extraction at short circuit condition. Static disorder can retain a
  stronger influence because it broadens the T-state energy distribution and
  modifies the population of low-energy states involved in non-radiative
  loss.

  We note here that the primary CT--CS scan herein varies $E_{CS}$
  at a fixed $E_{CT}$. Consequently, it does not change the CT3--T
  energy separation. A separate scan is therefore required to determine how
  changing the CT--CS alignment through $E_{CT}$ affects triplet
  recycling as will be discussed below.

  \subsubsection{LE-T offset}

  For the LE--T scan in Fig.\ref{offset}(c1)-(c6), $E_{LE}$,
  $E_{CT}$, and $E_{CS}$ were fixed, while the T energy was
  varied through $E_{T} = E_{LE} - \Delta{}E_{LE-T}$. The
  LE--T offset is equivalent to the singlet--triplet energy separation, i.e.
  $\Delta{}E_{ST}$. This offset is the singlet--triplet separation;
  a larger value stabilizes T relative to LE and CT3 and suppresses thermally
  activated T to CT3 return. Its device impact is strongest when CT3--CS
  separation is endothermic and triplet recycling is needed for efficient
  charge generation.

  Increasing $\Delta{}E_{LE-T}$ stabilizes the T state relative to
  LE and CT3. This favors triplet accumulation while suppressing thermally
  activated T-to-CT3 back transfer. Accordingly, IQE decreases and
  $V_{loss}$ increases with increasing LE--T offset, most strongly
  when CT3--CS separation is endothermic and recycling is required to
  generate free carriers.

  At the reference CT--CS offset of 0.10 eV, however, the LE--T dependence on
  dynamic disorder is weaker than the LE--CT and CT--CS dependences.
  Variations in $\lambda{}_{v}$ and $\lambda{}_{l}$
  produce only moderate changes in IQE, see Fig.\ref{offset}(c4,c5).
  Their influence is more visible in $V_{loss}$, see
  Fig.\ref{offset}(c1,c2), again showing that triplet kinetics affect
  open-circuit recombination more strongly than short-circuit charge
  extraction. Increasing $\sigma{}$ produces a larger response because
  disorder changes both the available triplet-state population and the
  distribution of recombination rates, as shown in
  Fig.\ref{offset}(c3,c6).

  The relatively weak LE--T dependence at the reference alignment does not
  imply that the T state is generally unimportant for charge generation.
  Instead, it suggests that the contribution of triplet recycling depends not
  only on the CT3-T energy separation, but also on the CT-CS alignment.

  \begin{figure*}[t]
\centering
\includegraphics[width=0.9\textwidth]{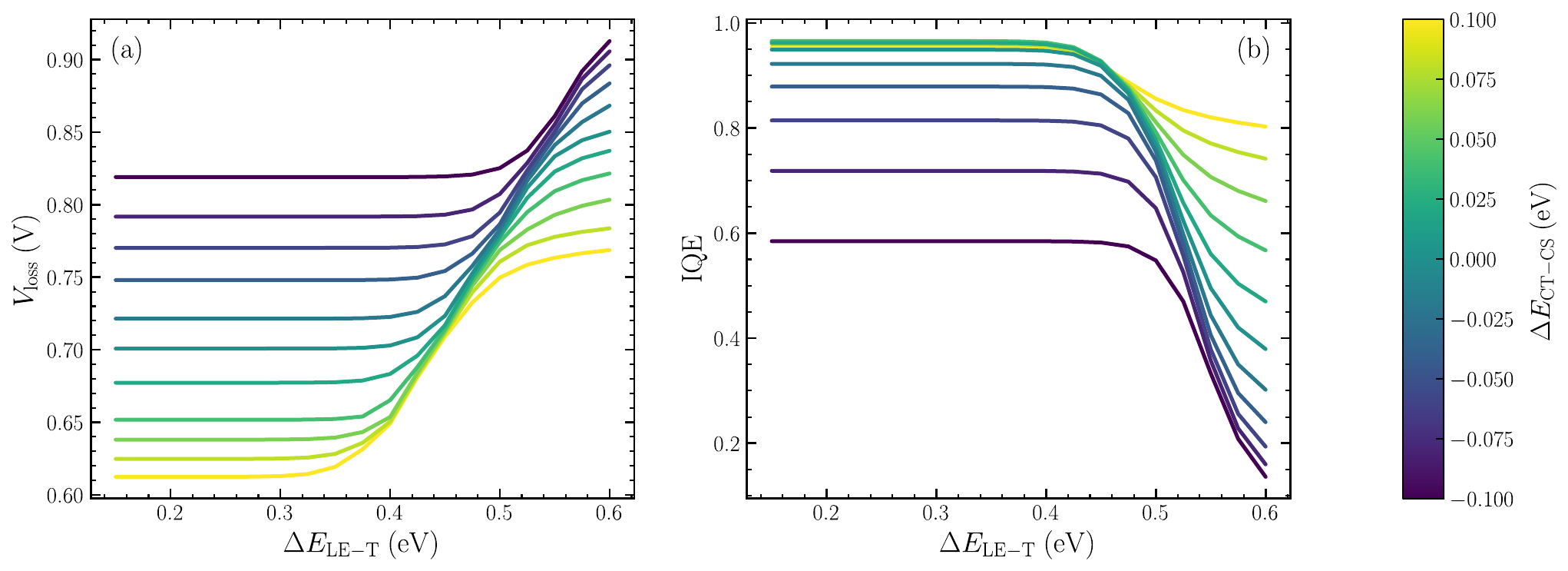}
\caption{Calculated $V_{loss}$ and IQE as functions of $\Delta{}E_{LE-T}$ for different CT--CS alignments. $E_{CS}$ was fixed while $E_{CT}$ was varied from $\Delta{}E_{CT-CS} = 0.1$ to $-0.1 eV$. Reducing the CT--CS offset increases the sensitivity to $\Delta{}E_{LE-T}$, demonstrating that triplet recycling is controlled by the coupled CT--CS and CT3--T energy alignment.}
\label{ct-cs}
\end{figure*}

  To further demonstrate the important role of CT-CS energy alignment, we
  performed the additional calculation shown in Fig.\ref{ct-cs}. Here,
  $E_{CS}$ was fixed and $E_{CT}$ was varied in the range
  of $-0.1 eV \leqslant{} \Delta{}E_{CT-CS}\leqslant{}  0.1 eV$.
  For each CT--CS alignment, $\Delta{}E_{LE-T}$ was varied from
  0.15 to 0.60 eV. Unlike the primary CT--CS scan, this protocol changes the
  CT3--T energy separation directly.

  At $\Delta{}E_{CT-CS} = 0.1 eV$, IQE decreases only from
  approximately 0.95 to 0.80 over the investigated LE--T range. When the
  CT--CS offset is reduced to 0 eV, IQE decreases from approximately 0.95 to
  0.38. At $\Delta{}E_{CT-CS} = -0.06 eV$, it decreases from
  approximately 0.81 to 0.19. Fig.\ref{ct-cs} shows that this change is
  continuous: the sensitivity to the LE--T offset increases progressively as
  $\Delta{}E_{CT-CS}$ becomes smaller.

  With $E_{CS}$ fixed, decreasing $\Delta{}E_{CT-CS}$
  lowers $E_{CT}$ and brings CT3 closer to the T state. The
  detailed-balance rate for back transfer is given by
  $k^{T\mathbin{\rightarrow}CT3}_{trans} =
  k^{CT3\mathbin{\rightarrow}T}_{trans}\frac{\left[CT3\right]_{0}}{\left[T\right]_{0}}$.
  Reducing the CT3--T separation increases the thermal accessibility of CT3
  from T. The recovered CT3 population can then dissociate into CS,
  completing the recycling pathway $T\mathbin{\rightarrow}
  CT3\mathbin{\rightarrow}CS$. For a large positive CT--CS offset, CT dissociation
  is already efficient, but the CT3--T separation suppresses back transfer
  from T. The T state then acts predominantly as a recombination loss
  channel. For a small or negative CT--CS offset, the CT3 state becomes
  thermally accessible from T and triplet recycling can contribute directly
  to the free-carrier yield. Therefore, neither $\Delta{}E_{LE-T}$
  nor $\Delta{}E_{CT-CS}$ alone determines triplet recycling.

  Overall, CT--CS energetics constitute an important bottleneck for
  free-carrier generation. The T state becomes important for charge
  generation when the CT3--T separation is sufficiently small to permit
  recycling, but it can influence voltage loss even when this condition is
  not satisfied. This distinction explains why triplet properties can have a
  strong effect on $V_{loss}$ while producing only a weak change in
  IQE under otherwise favorable charge-separation conditions.

  \subsection{Impact of different excited states}

  To separate the roles of the different excited states, we additionally
  varied the LE and CT state properties using the same ranges and EKDD as for
  the T state parameter scan. The corresponding results are shown in Appendix A
  and B as well as Fig.\ref{LE-offset} and Fig.\ref{CT-offset}.

  Although IQE and voltage loss are more strongly influenced by variations in
  LE and CT state parameters than by triplet-state properties, no single
  energetic offset alone can fully dictate a device's charge generation
  efficiency or voltage losses, consistent with the analysis above. The CT--CS
  offset is again important because it directly controls
  the conversion of the CT population into free carriers. At small or
  negative $\Delta{}E_{CT-CS}$, CT dissociation is weak or
  endothermic, allowing CT recombination, triplet formation, and reverse
  transfer to compete with charge separation. Consequently, the IQE becomes
  highly sensitive to $\lambda{}_{v}$, $\lambda{}_{l}$,
  and especially $\sigma{}$. Increasing the CT--CS driving force
  accelerates CT dissociation and causes the IQE curves to converge, but it
  also lowers $E_{CS}$ and therefore increases $V_{loss}$.
  Thus, the CT--CS alignment represents a central but previously overlooked
  trade-off between efficient charge generation and voltage loss. It also
  determines when triplet
  recycling becomes relevant: when the CT--CS driving force is weak, the
  CT3--T alignment can strongly influence recycling, whereas under favorable
  CT--CS alignment direct dissociation dominates and the influence of LE, CT,
  and T parameters on IQE is reduced.

  \begin{figure*}[t]
\centering
\includegraphics[width=0.9\textwidth]{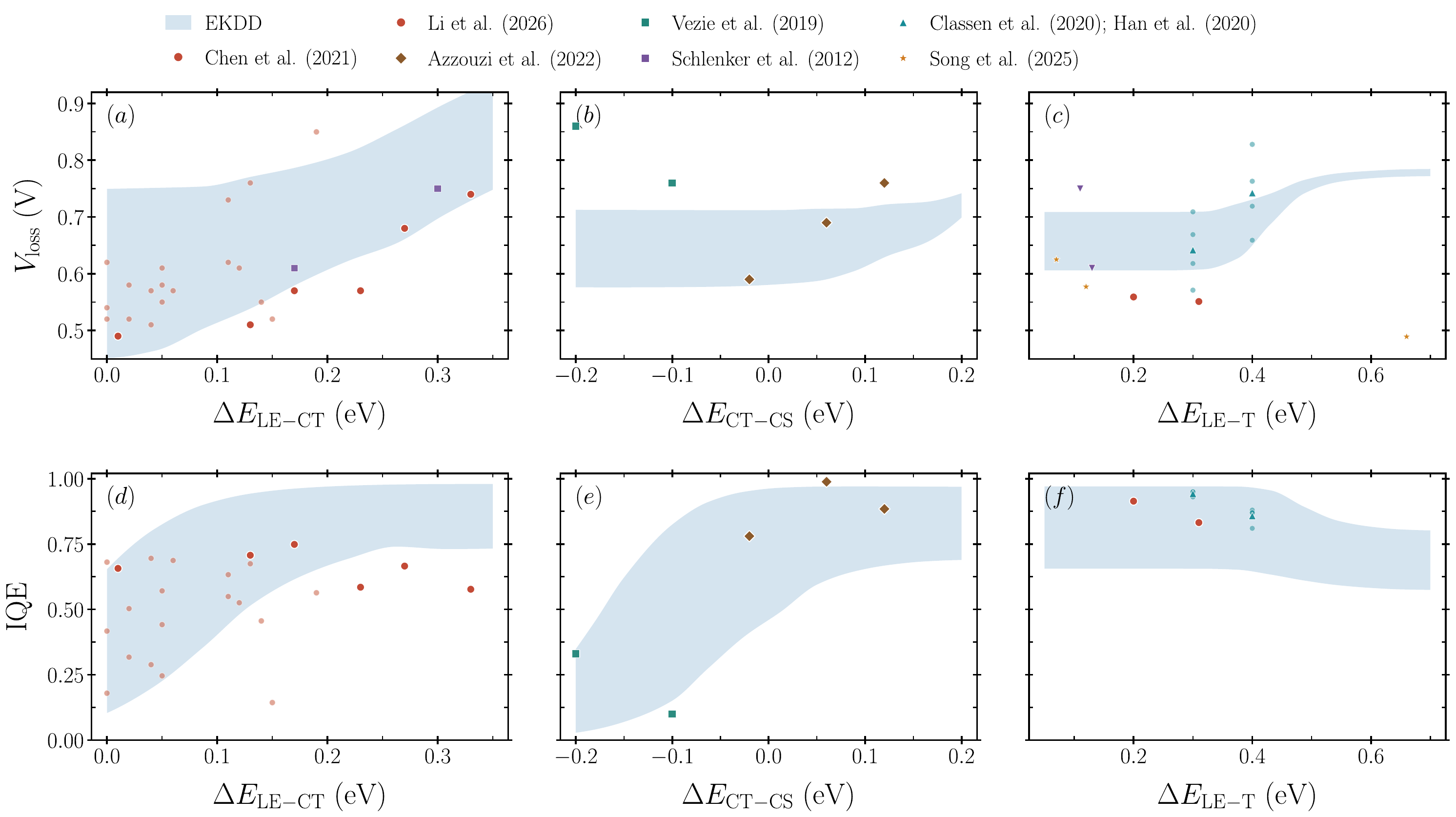}
\caption{Comparison of EKDD energy-offset trends with literature data. Calculated (a-c) voltage loss and (d-f) internal quantum efficiency as functions of the LE-CT, CT-CS, and LE-T energy offsets, respectively. The light-blue regions denote the comprehensive EKDD parameter envelope. Symbols show literature data from Chen et al. \cite{chen2021A}, Li et al. \cite{li2026recycling}, Azzouzi et al. \cite{azzouzi2022}, Vezie et al.\cite{vezie2019}, Schlenker et al. \cite{schlenker2012}, Classen et al. \cite{classen2020role}, Han et al. \cite{han2020}, and Song et al. \cite{song2025regulate}, as identified in the legend. The band quantifies model sensitivity to the scanned reorganization energies, oscillator strengths, and energetic disorder.}
\label{EKDD-comp}
\end{figure*}

  Using the method in Appendix: C, the comparison with reported data is shown
  in Fig.\ref{EKDD-comp}. The
  calculated parameter envelope overlaps a substantial fraction of the
  reported $V_{loss}$ and IQE values across the three offsets. The
  comparison supports the predicted increase in IQE with increasing LE--CT
  and CT--CS driving force, together with the accompanying increase in
  voltage loss. The LE--T data show larger scatter because the reported
  systems involve different recycling mechanisms and energetic definitions.
  Overall, the comparison indicates that the model captures the relevant
  energetic trends, while deviations reflect material-specific kinetics,
  uncertainty in experimentally inferred energy levels, and differences in
  IQE normalization.

  \section{Case study}

  Here we present two case studies to test the EKDD model and to connect its
  exciton kinetics with experimentally observed device behavior. Following
  Gillett et al. \cite{gillett2021role} and Li et al. \cite{li2026recycling},
  we first examine triplet-mediated recombination in PM6:Y6 and then
  triplet-assisted charge generation in D18:NFA blends. The first case tests
  whether the model reproduces a dominant terminal triplet recombination
  branch and its effect on voltage loss. The second tests whether the
  competition between CT3-to-T trapping and T-to-CT3 return explains
  material-dependent transient populations and calibrated steady-state
  current, voltage, and IQE. Together, the cases validate the model at both
  the state-population and device-performance levels and show why triplet
  states can act either as terminal loss channels or as recyclable
  intermediates, depending on energy alignment and transfer kinetics. 

  \subsection{Triplet-mediated recombination in PM6:Y6}

  To test whether EKDD reproduces the triplet-mediated recombination pathway
  identified by Gillett \emph{et al.} \cite{gillett2021role}, we used the
  reported PM6:Y6 parameters:  $k_{diss}^{CT3\mathbin{\rightarrow}CS}
  = 6.3\times{}10^{10} s^{-1}$, an inferred CT3-to-T
  back-charge-transfer rate $k^{CT3\mathbin{\rightarrow}T}_{trans} =
  k_{BCT} \sim 10^{11}-10^{12} s^{-1}$, $E_{LE} =
  1.6 eV$, $E_{CT} = 1.55 eV$, and $E_{T} = 1.2 eV$. Our
  model was solved under open-circuit conditions with constant 1-sun
  generation. Under open circuit condition, the photogenerated carriers
  cannot be extracted, so the steady-state carrier and exciton populations
  are determined by the competition between generation, CT dissociation, back
  transfer, and terminal recombination.

\begin{figure*}[t]
\centering
\includegraphics[width=0.9\textwidth]{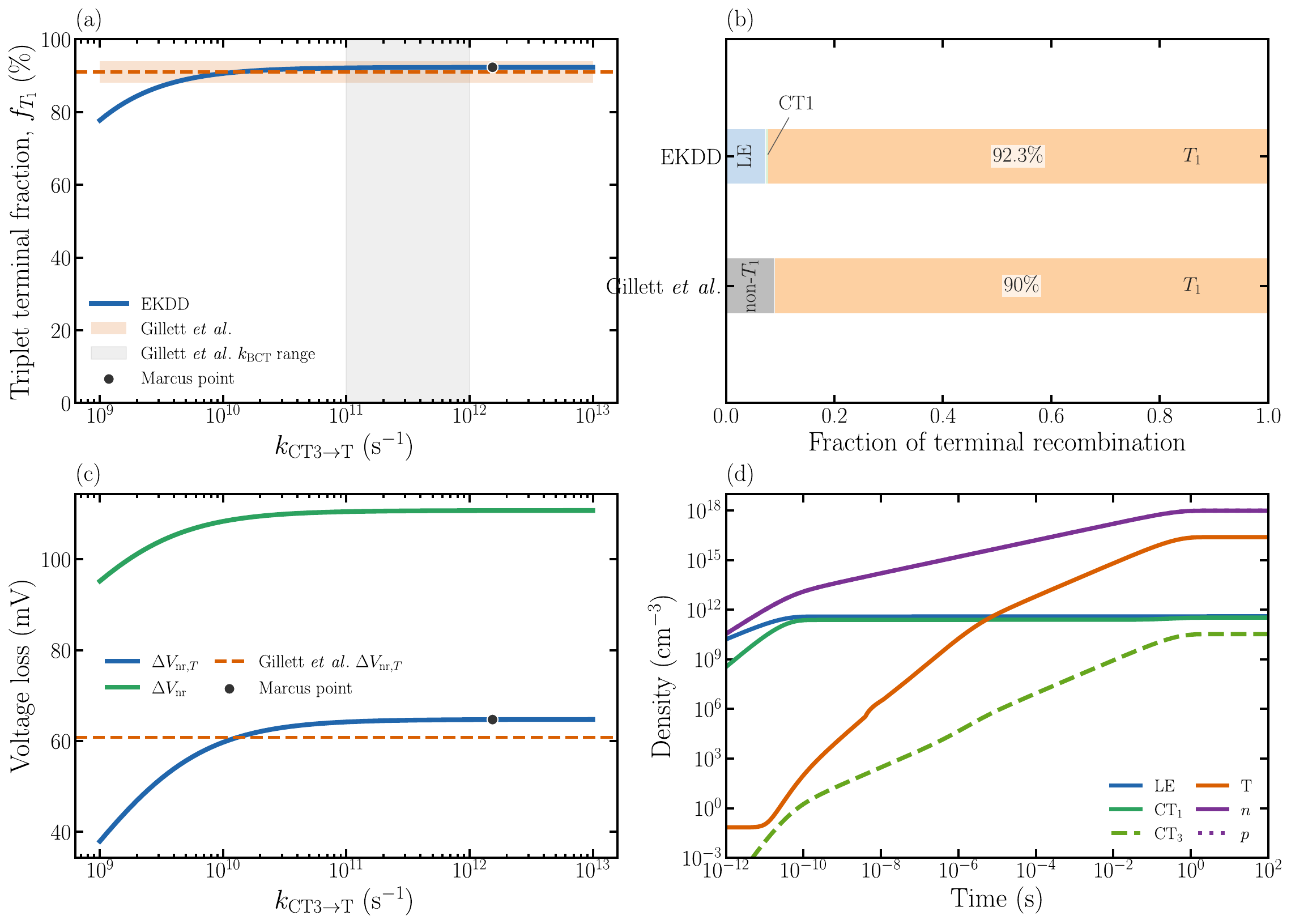}
\caption{Triplet-mediated recombination in PM6:Y6. (a) EKDD terminal triplet recombination fraction $f_{T}$ versus $k^{CT3\mathbin{\rightarrow}T}_{trans}$. The orange band denotes the 0.91$\pm{}$0.03 fraction reported from Gillett \emph{et al}.\cite{gillett2021role}, the grey band denotes the inferred $k_{BCT} = 10^{11}-10^{12} s^{-1}$ interval, and the marker denotes the Marcus rate calculated by EKDD. (b) Recombination partition at the Marcus point, showing the LE, CT1, and T contributions. (c) Triplet-only voltage penalty $\Delta{}V_{nr,T}$. (d) Transient population change at open-circuit condition of LE, CT1, CT3, T, electrons $n$, and holes $p$, calculated from 1 ps to 100 s. Solid and dashed curves distinguish the individual state populations, while the overlapping $n$ and $p$ curves reflect electron--hole symmetry in the homogeneous open-circuit diagnostic.}
\label{fraction}
\end{figure*}

  The triplet recombination fraction was defined from the terminal fluxes,

  \begin{align}
f_{T} = \frac{R_{T}}{R_{LE} + R_{CT1} + R_{ T}},
\end{align}

  where $R_{LE} = k_{rec, LE} \left[LE\right]$,
  $R_{CT1} = k_{rec,CT1} \left[CT1\right]$ and $R_{T}
  = k_{rec, T} \left[T\right]$. As shown in Fig.\ref{fraction}(a),
  $f_{T}$ increases from approximately 0.78 at
   $k^{CT3\mathbin{\rightarrow}T}_{trans} = 10^{9} s^{-1}$ to
  approximately 0.91 near $10^{10} s^{-1}$. It then approaches a
  plateau, remaining between 0.921 and 0.923 over the
  $10^{11}-10^{12} s^{-1}$ interval inferred from Gillett
  \emph{et al} \cite{gillett2021role}. At the Marcus point calculated by EKDD,
  $k^{CT3\mathbin{\rightarrow}T}_{trans} = 1.53\times{}10^{12}
  s^{-1}$, the model gives $f_{T} =0.923$. The remaining 7.7\% of
  terminal recombination is divided between the LE and CT1 channels, with
  7.35\% through LE and 0.36\% through CT1. The horizontal flux partition in
  Fig.\ref{fraction}(b) therefore reproduces the central conclusion of
  the paper: approximately 90\% of terminal free-carrier recombination
  proceeds through the non-emissive Y6 triplet state.

  The increase and subsequent saturation of $f_{T}$ can be
  understood from the competition at the CT3 state. At low
  $k^{CT3\mathbin{\rightarrow}T}_{trans}$, CT3 has sufficient time to
  dissociate into CS or undergo competing CT recombination before being
  transferred to T. Increasing $k^{CT3\mathbin{\rightarrow}T}_{trans}$
  redirects CT3 population into the triplet, thereby increasing the fraction
  of recombination that ultimately occurs through T. Once this transfer rate
  becomes comparable to or larger than the competing CT3 loss rates, most CT3
  population entering this kinetic branch is captured by T. Further
  increasing $k^{CT3\mathbin{\rightarrow}T}_{trans}$ therefore produces
  only a small change in $f_{T}$, giving rise to the plateau
  observed in Fig.\ref{fraction}(a). The detailed-balance reverse rate
  is also important for interpreting this result. For the PM6:Y6 energies,
  the equilibrium CT3 to T population ratio strongly favors the T state.
  Consequently, the forward CT3 to T transfer can be very rapid while the
  reverse T to CT3 rate remains much smaller. T therefore behaves primarily
  as a population reservoir and terminal recombination channel rather than as
  a rapidly reversible intermediate.

  The corresponding triplet-only non-radiative voltage penalty was calculated
  as

  \begin{align}
\Delta{}V_{nr,T} = -\frac{k_{B} T}{q}
    ln\left(1-f_{T}\right),
\end{align}

  here $R_{T}$ is non-radiative, and the total non-radiative voltage
  loss is

  \begin{align}
\Delta{}V_{nr} = -\frac{k_{B} T}{q}
    ln\left(\frac{R_{rad}}{R_{nr}+R_{rad}}\right).
\end{align}

  At the Marcus point, EKDD gives $f_{T}=0.923$; the T terminal
  non-radiative fraction is about 0.915, while LE and CT1 contribute about
  7.35\% and 0.36\%, respectively. Therefore, the 64.7 mV shown in
  Fig.\ref{fraction}(c) is a metric to identify the contribution of
  each recombination pathway, rather than an exact breakdown of the total
  device voltage loss. The result confirms the mechanism-level conclusion
  that the T branch dominates terminal recombination in PM6:Y6.

  The transient population calculation in Fig.\ref{fraction}(d)
  provides the complementary dynamical interpretation. Starting from thermal
  equilibrium (dark)  at open circuit condition, the LE and CT populations
  respond rapidly after the onset of constant illumination. LE to CT
  transition and CT dissociation occur in the sub-nanosecond timescale. The T
  population, in contrast, increases over many orders of magnitude and
  approaches its steady value only on much longer timescales, extending to
  approximately seconds.

  The much larger T population does not by itself determine the 90\%
  recombination fraction; rather, it reflects the long-lived nature of the
  triplet reservoir. The dominant triplet recombination fraction results from
  the combination of T accumulation and its terminal decay rate. The
  transient calculation shows that the system first populates LE, CT and CS
  state, after which the slowly evolving T reservoir controls the delayed
  recombination dynamics. This provides a mechanistic explanation for the
  strong sensitivity of $V_{oc}$ to triplet kinetics under
  open-circuit conditions.

  The agreement between $f_{T} = 0.923$ from EKDD and the
  approximately 90\% fraction reported by Gillett \emph{et al.}
  \cite{gillett2021role} should be interpreted as a mechanism-level
  validation. Gillett \emph{et al.} obtained parameter $alpha$ by
  fitting a reduced transient-absorption model, whereas EKDD resolves LE,
  CT1, CT3, T, and free-carrier populations explicitly. The present
  calculation therefore does not imply that all device-level voltage losses
  are reproduced by the triplet branch alone. Instead, it demonstrates that
  the same dominant pathway emerges from the microscopic five-state rate
  model: rapid CT3 to T transfer, slow triplet depletion, and terminal
  recombination mainly through the T state.

  \subsection{Triplet-exciton recycling in D18:NFA blends}

  To test whether our EKDD framework captures the triplet-recycling mechanism
  reported in Ref.\cite{li2026recycling}, we calculated the free charge and
  triplet exciton population change for the D18:BrQx-4BO and D18:FTh-4F
  systems.

  The reference calculation follows the four-state kinetic scheme used in
  Supplementary Note 6 of Li \emph{et al.} \cite{li2026recycling}, and
  parameters follows Table 3. Different from Ref.\cite{langentepe2026dark},
  we here populate LE first instead of CS. The relative importance of triplet
  recycling is characterized by

  \begin{align}
K = k_{trans}^{T\mathbin{\rightarrow}CT3}/k_{trans}^{CT3\mathbin{\rightarrow}T},
\end{align}

  a Gaussian LE-generation pulse was used to reproduce the population
  initialization in Ref.\cite{li2026recycling} ns-TAS analysis:

  \begin{align}
G\left(t\right) = G_{0} \exp \left[-\left(t-t_{0}\right)^{2}/2
    \sigma{}_{t}^{2}\right]
\end{align}

  with a total generated LE population of approximately
  $5.2\times{}10^{17} cm^{-3}$, a pulse FWHM of 1.32 ns, and a
  pulse centre at 0.66 ns.
\begin{table*}[t]
\caption{Parameters used for the density comparison of D18:BrQx-4BO and D18:FTh-4F.}
\centering
\footnotesize
\setlength{\tabcolsep}{5pt}
\renewcommand{\arraystretch}{1.14}
\begin{tabular}{llll}
\toprule

      \emph{Parameter}
     & 
      D18:BrQx-4BO
     & 
      D18:FTh-4F
     & 
      \emph{Unit}
     \\
\midrule

      $E_{LE}$
     & 
      1.49
     & 
      1.41
     & 
      $eV$
     \\

      $E_{CT}$
     & 
      1.46
     & 
      1.38
     & 
      $eV$
     \\

      $E_{T}$
     & 
      1.18
     & 
      1.21
     & 
      $eV$
     \\

      $\Delta{}E_{ST} = E_{LE} - E_{T}$
     & 
      0.31
     & 
      0.20
     & 
      $eV$
     \\

      $E_{CS}$
     & 
      1.20
     & 
      1.20
     & 
      $eV$
     \\

      Initial LE population
     & 
      $5.20\times{}10^{17}$
     & 
      $5.19\times{}10^{17}$
     & 
      $\mathrm{cm}^{-3}$
     \\

      $k_{diss}^{CT1\mathbin{\rightarrow}CS}$
     & 
      $1.0\times{}10^{11}$
     & 
      $1.0\times{}10^{11}$
     & 
      $s^{-1}$
     \\

      $k_{diss}^{CT3\mathbin{\rightarrow}CS}$
     & 
      $1.0\times{}10^{11}$
     & 
      $1.0\times{}10^{11}$
     & 
      $s^{-1}$
     \\

      $k_{rec,CT1}$
     & 
      $1.0\times{}10^{9}$
     & 
      $1.0\times{}10^{9}$
     & 
      $s^{-1}$
     \\

      $k_{rec,T}$
     & 
      $1.0\times{}10^{6}$
     & 
      $1.0\times{}10^{6}$
     & 
      $s^{-1}$
     \\

      $k_{trans}^{CT3\mathbin{\rightarrow}T}$
     & 
      $8.5\times{}10^{10}$
     & 
      $2.2\times{}10^{10}$
     & 
      $s^{-1}$
     \\

      $k_{trans}^{T\mathbin{\rightarrow}CT3}$
     & 
      $1.3\times{}10^{8}$
     & 
      $3.9\times{}10^{9}$
     & 
      $s^{-1}$
     \\

      $B_{for}^{CS\mathbin{\rightarrow}CT}$
     & 
      $1.7\times{}10^{-9}$
     & 
      $1.7\times{}10^{-9}$
     & 
      $\mathrm{cm}^{3}\,\mathrm{s}^{-1}$
     \\

      Active-layer thickness
     & 
      120
     & 
      120
     & 
      $\mathrm{nm}$
     \\
\bottomrule
\end{tabular}
\end{table*}

\begin{figure*}[t]
\centering
\includegraphics[width=0.9\textwidth]{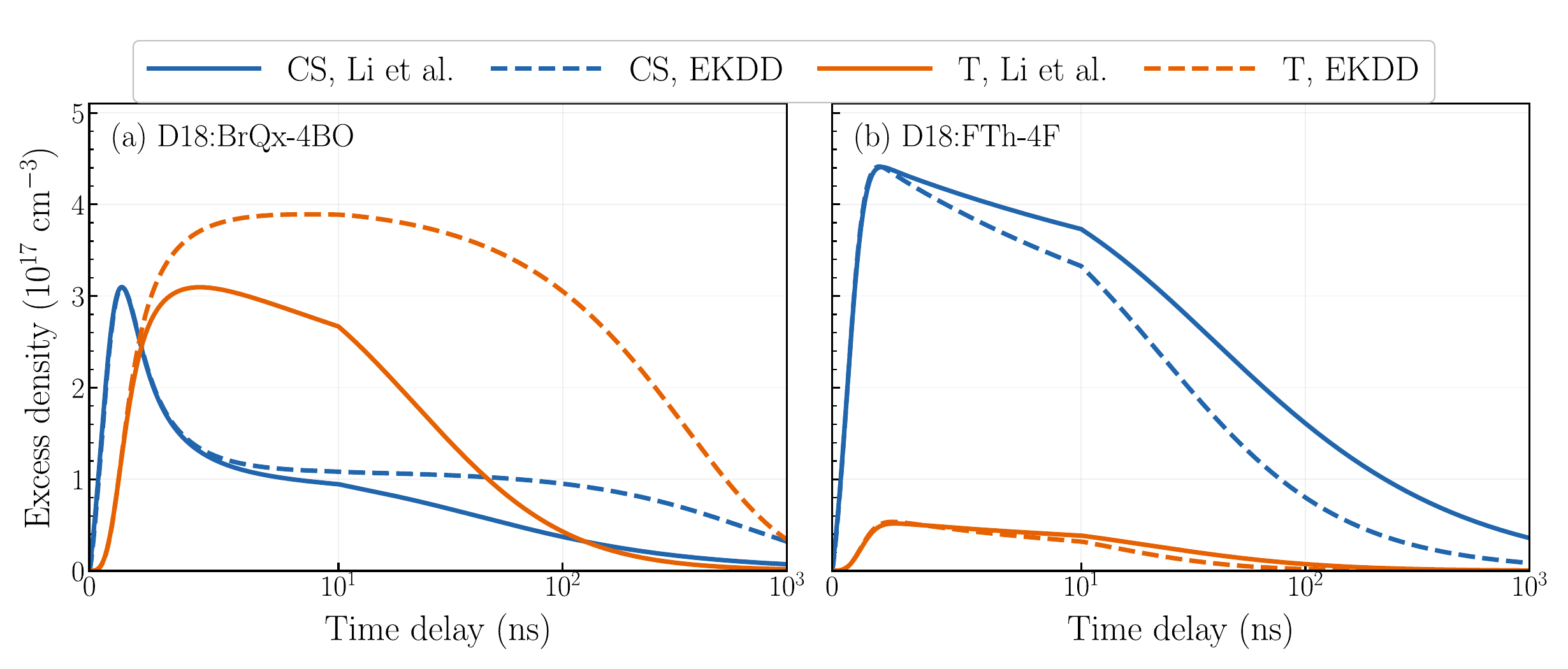}
\caption{\textbf{Calculated CS and T population densities for D18:BrQx-4BO and D18:FTh-4F.} CS is shown in blue and T in orange in both calculations; solid lines denote the paper calculation and dashed lines denote our EKDD calculation.}
\label{density}
\end{figure*}

  Fig.\ref{density} shows the resulting CS and T population. Both
  calculations reproduce the material-dependent trend reported by Li et al.,
  D18:BrQx-4BO undergoes rapid CT3-to-T conversion and substantial triplet
  accumulation, whereas D18:FTh-4F retains a larger CS population and
  exhibits less persistent triplet trapping. The remaining difference in the
  BrQx-4BO amplitude is expected because the authors directly initializes CS
  states, while EKDD includes LE loss, LE to CT transfer, CT dissociation,
  carrier transport, and density-dependent CT3 formation. Therefore, this
  comparison validates the kinetic ordering and recycling mechanism, rather
  than constituting a fitted transient-absorption spectrum.

\begin{figure*}[t]
\centering
\includegraphics[width=0.95\textwidth]{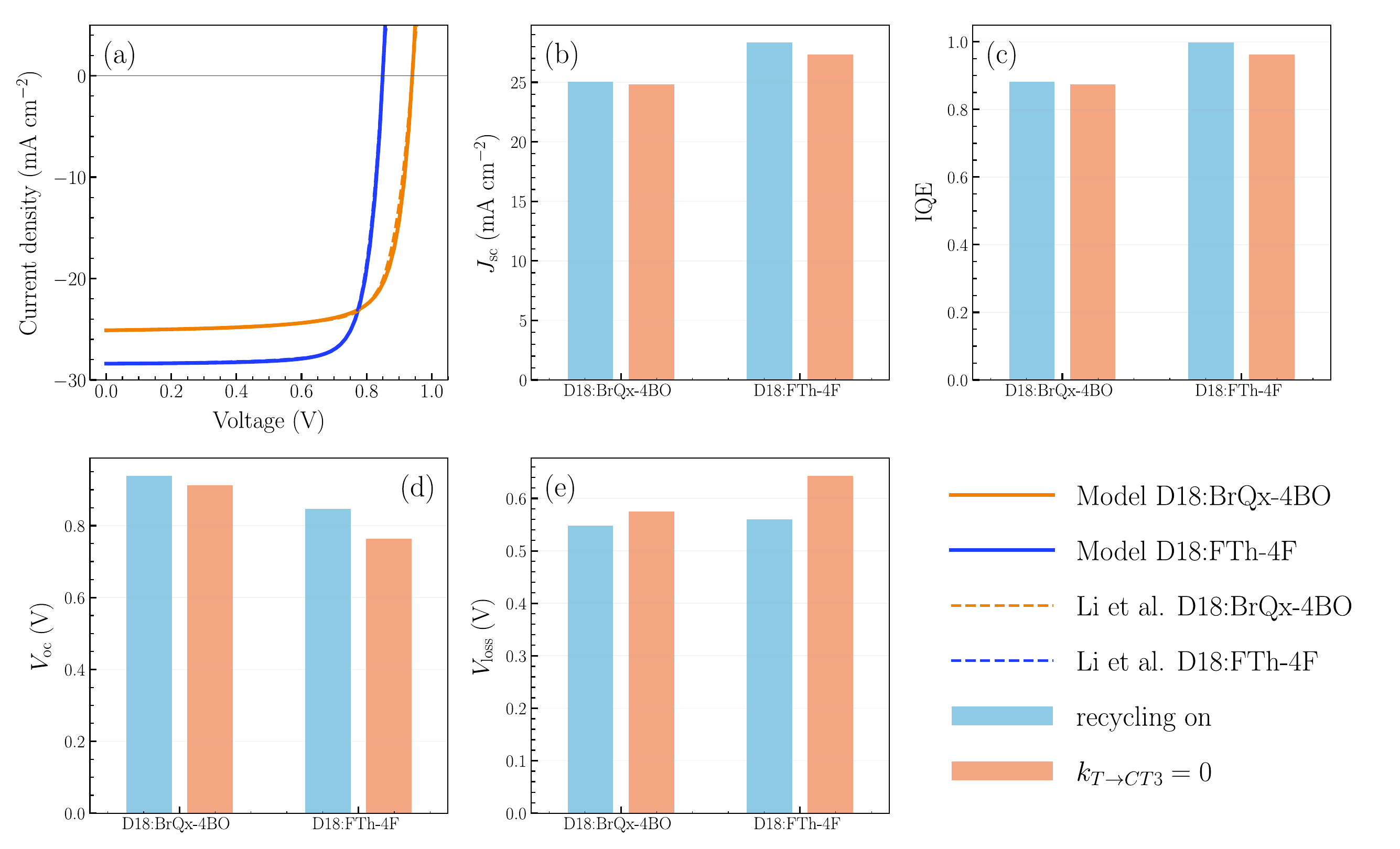}
\caption{\textbf{Steady-state effect of triplet recycling in D18:BrQx-4BO and D18:FTh-4F.} (a) J-V curve, (b) $J_{sc}$, (c) IQE, (d) $V_{oc}$, and (e) $V_{loss}$. “Recycling off” sets only $k_{ trans}^{T\mathbin{\rightarrow}CT3}=0$, while retaining CT3 to T transfer and triplet recombination. Efficient recycling in FTh-4F increases charge collection and reduces voltage loss substantially, whereas the effect is weak in BrQx-4BO.}
\label{dv}
\end{figure*}

  We next quantified the device-level responses by comparing steady-state
  simulations with active recycling or inactive recycling ($k_{
  trans}^{T\mathbin{\rightarrow}CT3}=0$), the comparison was calibrated via
  fitting the J-V curve, as shown in Fig.\ref{dv}. Importantly,
  $k_{ trans}^{T\mathbin{\rightarrow}CT3}$ and triplet recombination
  remain active in the latter calculation; this is a loss-of-recycling
  control, not removal of the triplet state.

  For D18:BrQx-4BO, activating recycling increases $J_{sc}$ only
  from 24.86 to 25.10 $mA/cm^{2}$ (0.965\%), corresponding to an IQE
  increase from 0.8755 to 0.8840. The $V_{oc}$ increases by 26.4 mV,
  from 0.9142 to 0.9406 V, reducing $V_{loss}$ from 0.5758 to 0.5494
  V. Its small $K$ value therefore makes triplet recovery ineffective:
  most triplets remain trapped or recombine before returning to CT3.

  D18:FTh-4F benefits a bit more from recycling. Restoring $k_{
  trans}^{T\mathbin{\rightarrow}CT3}$ increases $J_{sc}$ from 27.384 to
  28.393 $mA/cm^{2}$ (3.685\%), and raises IQE from 0.9642 to 0.9998.
  Simultaneously, $V_{oc}$ rises from 0.7657 to 0.8488 V, reducing
  $V_{loss}$ from 0.6443 to 0.5612 V. Recycling therefore improves
  both carrier collection and voltage retention in this calibrated case
  because returned CT3 population can dissociate before being lost.

  Although both materials have positive CT-CS offsets, this does not make
  triplet recycling irrelevant. Its importance is determined by the complete
  competition $T\mathbin{\rightarrow}CT3\mathbin{\rightarrow}CS$ against triplet decay
  and re-trapping. The substantially larger $K$ in D18:FTh-4F enables
  triplets to re-enter the dissociative CT3 channel, whereas the small
  $K$ in D18:BrQx-4BO leads to terminal triplet loss. These results
  support the central conclusion of Li et al.: reducing the energetic and
  kinetic barrier for triplet recovery can convert T from a recombination
  reservoir into a recyclable intermediate that supports photocurrent and,
  for the present calibrated parameters, preserves part of the voltage. The
  outcome is controlled by the complete T--CT3--CS competition rather than by
  the presence of a positive CT--CS offset alone.

  Overall, triplet recycling enhances current generation and open circuit
  voltage, but the degree will depend on the energetics. Importantly, the
  impact on $J_{sc}$ enhancement is relatively small for FTh-4F
  (3.6\%). Besides, the difference in optical gap has already accounted for
  major part of the current density difference between D18:FTh-4F and
  D18:BrQx-4BO without considering triplet recycling (27 vs. 25 mA/cm2),
  therefore, we consider the impact on current gain is very small with
  triplet recycling in the systems studied herein. Nevertheless, triplet
  recycling can still be considered as a pathway for more efficient charge
  generation and reduced voltage losses as will be discussed later.

  \section{Discussion}

  \subsection{Charge extraction with triplet recycling}

  When using transition rate constants determined typically from thin film
  TAS that are equivalent to open circuit conditions, it remains unclear
  whether changes in transition rates in thin film TAS can lead to large
  variations in transient and steady state current. To determine how the
  kinetic rates obtained from open-circuit transient-absorption spectroscopy
  (TAS) affect charge extraction, we used the TAS-derived rates as intrinsic
  local rate constants in EKDD under short circuit conditions. The electrical
  boundary condition was held at $V_{app}=0$ with Ohmic extracting
  contacts, and the transfer and recombination rates were not changed. The
  current response is then obtained from the resulting electron and hole
  transport currents together with the displacement current. The extracted
  charge was evaluated as

  \begin{align}
Q_{ext} = \int _{0}^{t_{f}}\Delta{}J_{ext}\left(t\right)
    d t,
\end{align}

  and it can be normalized to the generated pulse charge, we defined as the
  transient charge collection fraction

  \begin{align}
\eta{} = \frac{\int _{0}^{t_{f}}\Delta{}
    J_{ext}\left(t\right)d t}{q  N_{abs} L}.
\end{align}

  Unlike IQE defined in Eq.(\ref{iqe}), the low $\eta{}$
  is expected because most of the initially LE excitation is just held on
  several nanosecond.

\begin{figure*}[t]
\centering
\includegraphics[width=0.9\textwidth]{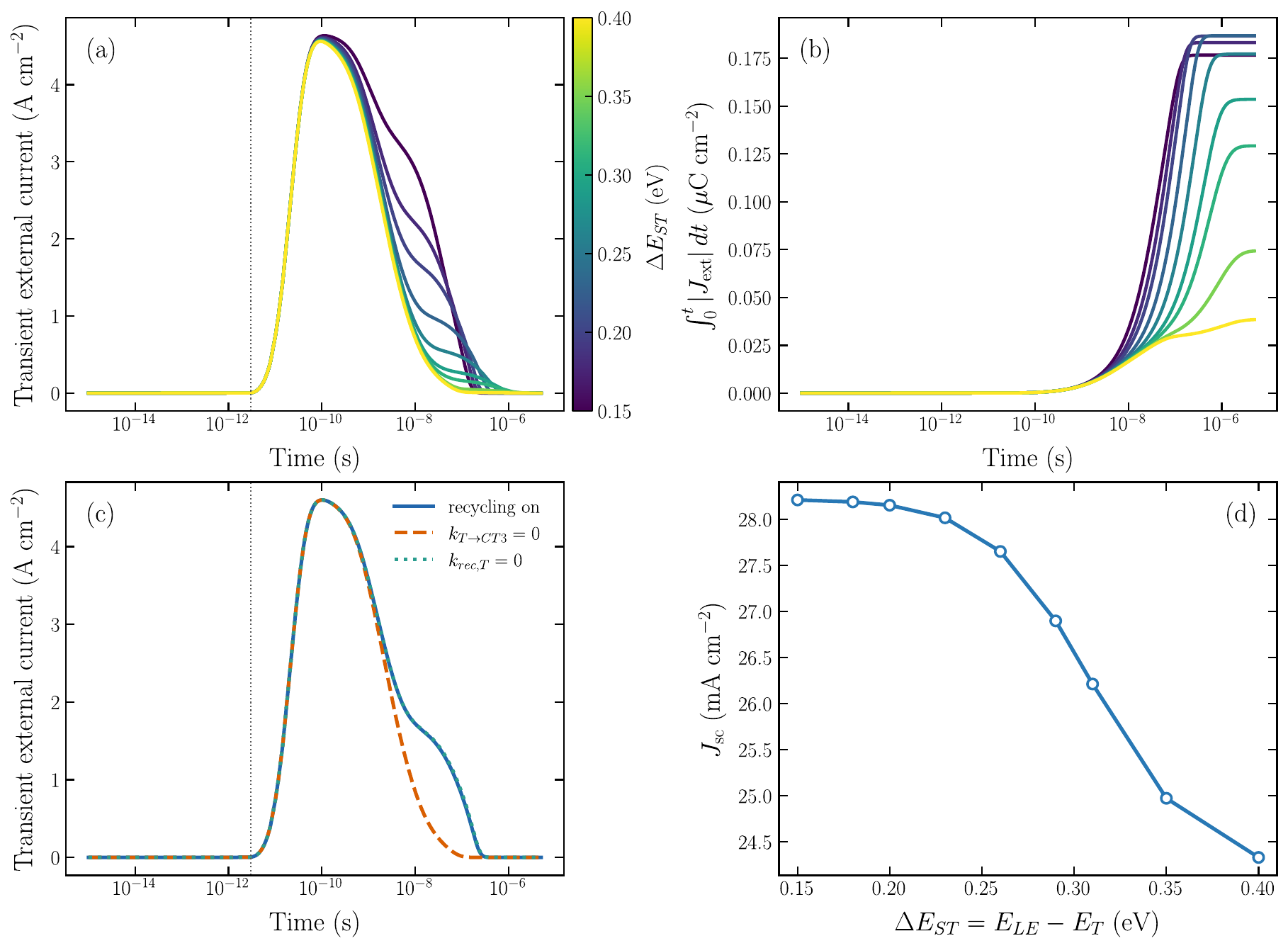}
\caption{Short-circuit charge extraction in the D18:FTh-4F device. (a) Transient external current; line colour denotes $\Delta{}E_{ST}$, and the dotted line marks the pulse centre. (b) Cumulative integrated external current, $\int _{0}^{t}\Delta{} J_{ext}\left(t\right)d t$, for the same offsets. (c) Recycling controls at $\Delta{}E_{ST}=0.2$ eV: the complete model, $k_{trans}^{T\mathbin{\rightarrow}CT3}=0$ with CT3 $\mathbin{\rightarrow}$ T retained, and $k_{rec,T}=0$ with both transfer directions retained. (d) Steady-state one-sun $J_{sc}$ from independent EKDD J--V solutions.}
\label{collection}
\end{figure*}

  We simulated the D18:FTh-4F device after a spatially uniform 532-nm pulse
  with a 1-ps full width at half maximum and a centre time of 3 ps. The
  excitation was applied only to the LE state, generated approximately
  $N_{abs} = 6.0\times{}10^{17} cm^{-3}$ excitations.
  $E_{LE} = 1.41 eV$, $E_{CT} = 1.38 eV$ and
  $E_{CS} = 1.20 eV$ were fixed, while $\Delta{}E_{ST} =
  E_{LE} - E_{T}$ was varied from 0.15 to 0.40 eV. The
  CT3$\mathbin{\rightarrow}$T Marcus rate, the detailed-balance
  T$\mathbin{\rightarrow}$CT3 rate, and $k_{rec, T}$ were recalculated
  for each offset. The four resulting observables are summarized in
  Fig.\ref{collection}.

  In Fig.\ref{collection}(a), the prompt short-circuit current peak
  changes only from approximately 4.635 to 4.561 $A/cm^{2}$ over the
  complete offset range. This weak dependence occurs because the leading peak
  is dominated by prompt LE to CT to CS conversion and the
  displacement-current response to the 1-ps perturbation. It occurs before
  the slower CT3--T branch has redistributed a significant fraction of the
  population. The several $A/cm^{2}$  amplitude is therefore an
  instantaneous transient response and should not be interpreted as a
  steady-state photocurrent.

  The influence of $\Delta{}E_{ST}$ appears in the delayed current
  tail in Fig.\ref{collection}(a) and in the cumulative current in
  Fig.\ref{collection}(b). Increasing $\Delta{}E_{ST}$ from
  0.15 to 0.40 eV increases $k_{trans}^{CT3\mathbin{\rightarrow}T}$
  from $1.45\times{}10^{11}$ to $1.71\times{}10^{12}
  s^{-1}$, while the detailed-balance
  $k_{trans}^{T\mathbin{\rightarrow}CT3}$ decreases from
  $1.4\times{}10^{9}$ to $1.04\times{}10^{6}$.
  Consequently, $K = k_{trans}^{T\mathbin{\rightarrow}CT3}/k_{trans}^{CT3\mathbin{\rightarrow}T}$
  decreases by more than four orders of magnitude. Triplet formation becomes
  more favorable, but return to CT3 becomes progressively less probable. The
  population therefore accumulates in T and is increasingly lost through
  terminal triplet recombination rather than recycled into the
  CT3$\mathbin{\rightarrow}$CS extraction pathway.

  The integrated external charge, $Q_{ext}$ varies from 0.038 to
  0.187 $\mu{}C/cm^{2}$, corresponding to a collected fraction of
  6.4--31.1\% of the generated pulse charge. The maximum collection occurs
  near $\Delta{}E_{ST} = 0.2-0.23 eV$. At small offsets, T can act
  as a temporary reservoir because the T$\mathbin{\rightarrow}$CT3 return
  remains sufficiently fast to support delayed extraction. At larger offsets,
  the reverse rate becomes too slow, and the accumulated triplet population
  is predominantly removed by $k_{rec,T}$. The decay tail therefore
  contracts and the integrated charge decreases to only 6.4\% at
  $\Delta{}E_{ST} = 0.4$ eV.

  The control simulations in Fig.\ref{collection}(c) isolate the role
  of recycling at $\Delta{}E_{ST} = 0.2$ eV. With both transfer
  directions and triplet recombination active, the collected fraction is
  31.1\%. Setting only $k_{trans}^{T\mathbin{\rightarrow}CT3}=0$, while
  retaining CT3$\mathbin{\rightarrow}$T formation, reduces the collected
  fraction to 5.8\%, an approximately 81\% relative loss. This control
  preserves triplet formation but prevents the triplet population from
  returning to the dissociable CT3 state. Conversely, setting only
  $k_{rec,T}=0$ while retaining both transfer directions increases
  the collected fraction slightly to 32.2\%. The three traces have nearly
  identical prompt peaks but diverge strongly in the delayed tail,
  demonstrating that the TAS-derived reverse transfer rate controls the
  additional charge extracted after the initial photogeneration event.

  The device-scale response is shown in Fig.\ref{collection}(d). The
  steady-state one-sun short-circuit current decreases from 28.21 to 24.33
  $mA/cm^{2}$ as $\Delta{}E_{ST}$ increases from 0.15 to
  0.40 eV, corresponding to a 13.7\% reduction. The steady-state dependence is
  smoother than the pulsed integrated response because it reflects a
  continuously maintained balance among generation, transport, electrostatics
  and recombination. Nevertheless, both observables show the same physical
  trend: increasing $\Delta{}E_{ST}$ reduces the recycling ratio
  and lowers charge extraction.

  Taken together, Fig. \ref{collection}(a--d) establishes how
  open-circuit TAS kinetics can be tested under short-circuit conditions. The
  TAS rates do not substantially modify the prompt current peak, but they
  determine the charge extraction time, the integrated transient charge and
  the steady-state $J_{sc}$. Efficient T$\mathbin{\rightarrow}$CT3
  recycling preserves the productive CT3$\mathbin{\rightarrow}$CS pathway,
  whereas suppression of this return channel converts triplet formation into
  an extraction loss.

  \subsection{Design rules}

  To establish energetic design rules, we considered the two connected state
  sequences that control charge generation: T--CT3--CS and LE--CT--CS.
  Two-dimensional steady-state EKDD scans were performed using the reference
  parameters in Table 2. In the T--CT3--CS scan, $E_{LE}=1.50$ eV
  and $E_{CT}=1.30$ eV were fixed. The CT--CS offset was varied
  through $E_{CS}=E_{CT}-\Delta{}E_{CT-CS}$, and the
  T-state energy was varied independently according to
  $E_{T}=E_{CT}- \Delta{}E_{CT-T}$. This protocol separates
  the thermodynamic driving force for CT3 dissociation from the energetic
  accessibility of the T to CT3 recycling step.

  In the LE--CT--CS scan, $E_{LE}=1.50$ eV and $E_{T}=1.15$
  eV were fixed. The CT energy was changed according to
  $E_{CT}=E_{LE}-\Delta{}E_{LE-CT}$, and the CS energy was
  then set by $E_{CS}=E_{CT}-\Delta{}E_{CT-CS}$. Unlike the
  first scan, changing $E_{CT}$ here also changes the CT--T offset.
  The second map therefore captures the coupled effects of LE-to-CT transfer,
  CT3 dissociation, voltage retention, and triplet recycling along a
  material-level LE--CT--CS energy sequence. At every point in both maps, the
  Marcus forward-transfer rates, detailed-balance reverse rates, FCWD
  recombination constants, equilibrium state densities, and
  $B_{for}^{CS\mathbin{\rightarrow}CT}$ were recalculated. The
  transport gap was changed consistently with $E_{CS}$, nine
  distributed states per excited state and $\sigma{}=0.01$ eV were
  used, and the negligible T radiative channel was omitted.

  \begin{figure*}[t]
\centering
\includegraphics[width=0.91\textwidth]{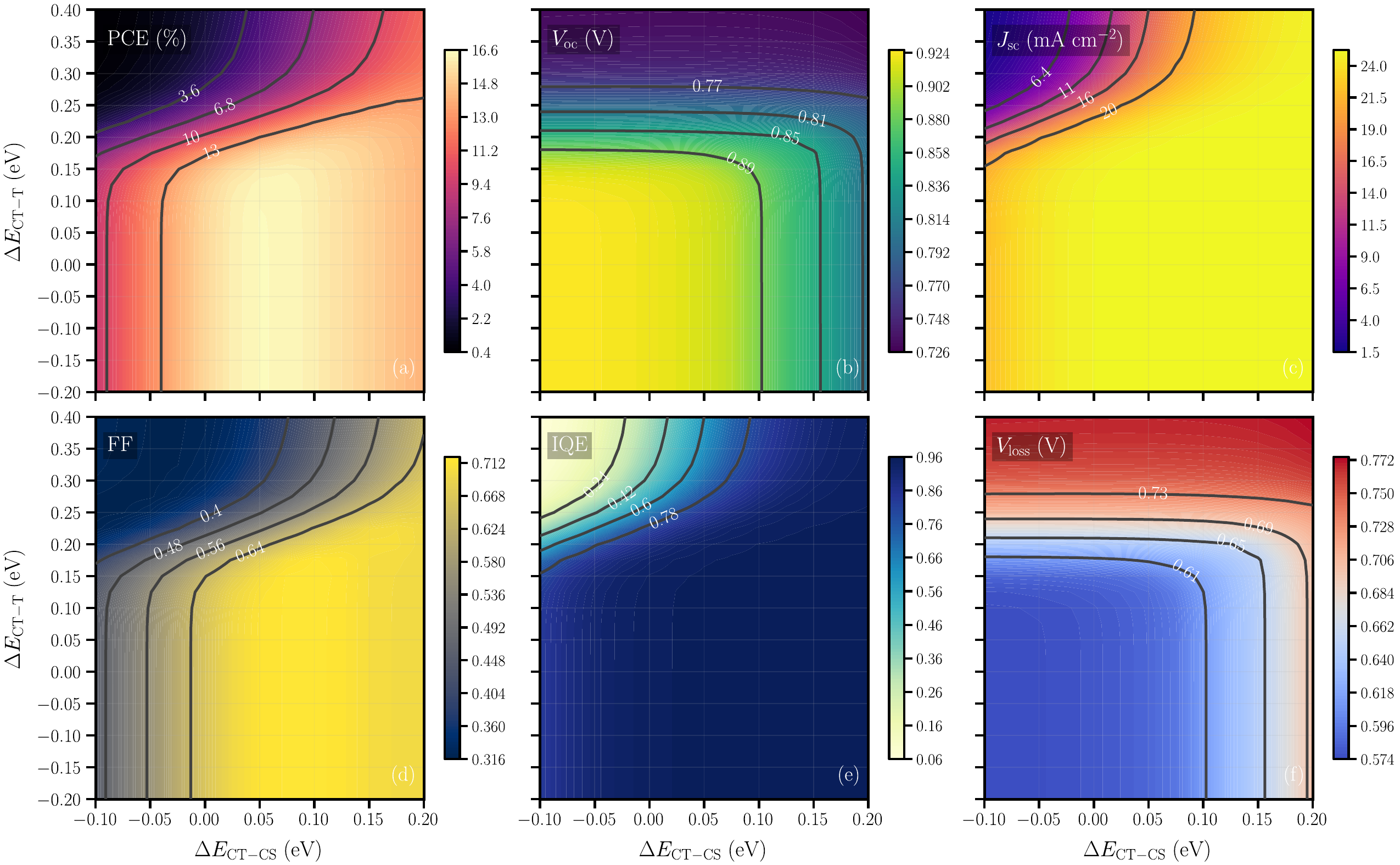}
\caption{\textbf{Two-dimensional EKDD design map for triplet-exciton recycling.} The horizontal axis is $\Delta{}E_{CT-CS}$, and the vertical axis is $\Delta{}E_{CT-T}$. Panels show (a) PCE, (b) $V_{oc}$, (c) $J_{sc}$, (d) FF, (e) IQE, and (f) $V_{loss}$. Color shading represents the calculated metric values, and contour labels indicate representative numerical levels.}
\label{T-CT-CS}
\end{figure*}

  Fig.\ref{T-CT-CS} shows that the CT--CS and CT--T offsets play
  distinct but coupled roles. From the definition
  $E_{CS}=E_{CT}-\Delta{}E_{CT-CS}$, a positive CT--CS
  offset places CS below CT3 and makes CT3-to-CS dissociation exothermic. IQE
  and $J_{sc}$ therefore rise rapidly and then approach a plateau as
  the offset becomes positive [panels (c) and (e)]. A negative offset makes
  dissociation endothermic, and charge generation then becomes much more
  sensitive to the CT--T alignment. By contrast, increasing the positive
  CT--CS offset lowers $E_{CS}$, reducing the attainable
  quasi-Fermi-level splitting. Consequently, $V_{oc}$ decreases and
  $V_{loss}$ increases [panels (b) and (f)]. CT--CS energetics thus
  impose the primary trade-off between charge-separation efficiency and
  voltage retention.

  The CT--T offset determines whether population trapped in T can return to
  dissociable CT3. This kinetic accessibility is summarized by
  $K=k_{trans}^{T\mathbin{\rightarrow}CT3}/k_{trans}^{CT3\mathbin{\rightarrow}T}$.
  For the present parameters, $K\approx{}1$ at
  $\Delta{}E_{CT-T}=0$ eV, whereas it decreases to
  $2.1\times{}10^{-2}$, $3.0\times{}10^{-3}$, and
  $4.4\times{}10^{-4}$ at 0.10, 0.15, and 0.20 eV, respectively.
  The influence of this offset is strongest when CT3-to-CS dissociation is
  endothermic or only weakly exothermic: as $K$ decreases, T becomes a
  terminal reservoir, lowering IQE, $J_{sc}$, FF, and PCE. When the
  CT--CS offset is strongly positive, rapid direct dissociation preserves
  much of the current and partly hides inefficient recycling, although it
  does not recover the voltage lost through triplet formation and
  recombination.

  The combined response produces the PCE ridge in Fig.\ref{T-CT-CS}(a).
  The maximum PCE is 16.42\% at $\Delta{}E_{CT-CS}=0.05$ eV and
  $\Delta{}E_{CT-T}=-0.20$ eV, with $V_{oc}=0.914$ V,
  $J_{sc}=25.18 mA/cm^{-2}$, FF = 0.713, and IQE = 0.953. The
  region within 98\% of this maximum spans
  $\Delta{}E_{CT-CS}=0.05-0.075$ eV and
  $\Delta{}E_{CT-T}=-0.20-0.15$ eV. The broad CT--T tolerance in
  this high-PCE region arises because the small positive CT--CS offset
  already provides efficient direct dissociation; under endothermic CT--CS
  conditions, the acceptable CT--T range would be substantially narrower.

  \begin{figure*}[t]
\centering
\includegraphics[width=0.91\textwidth]{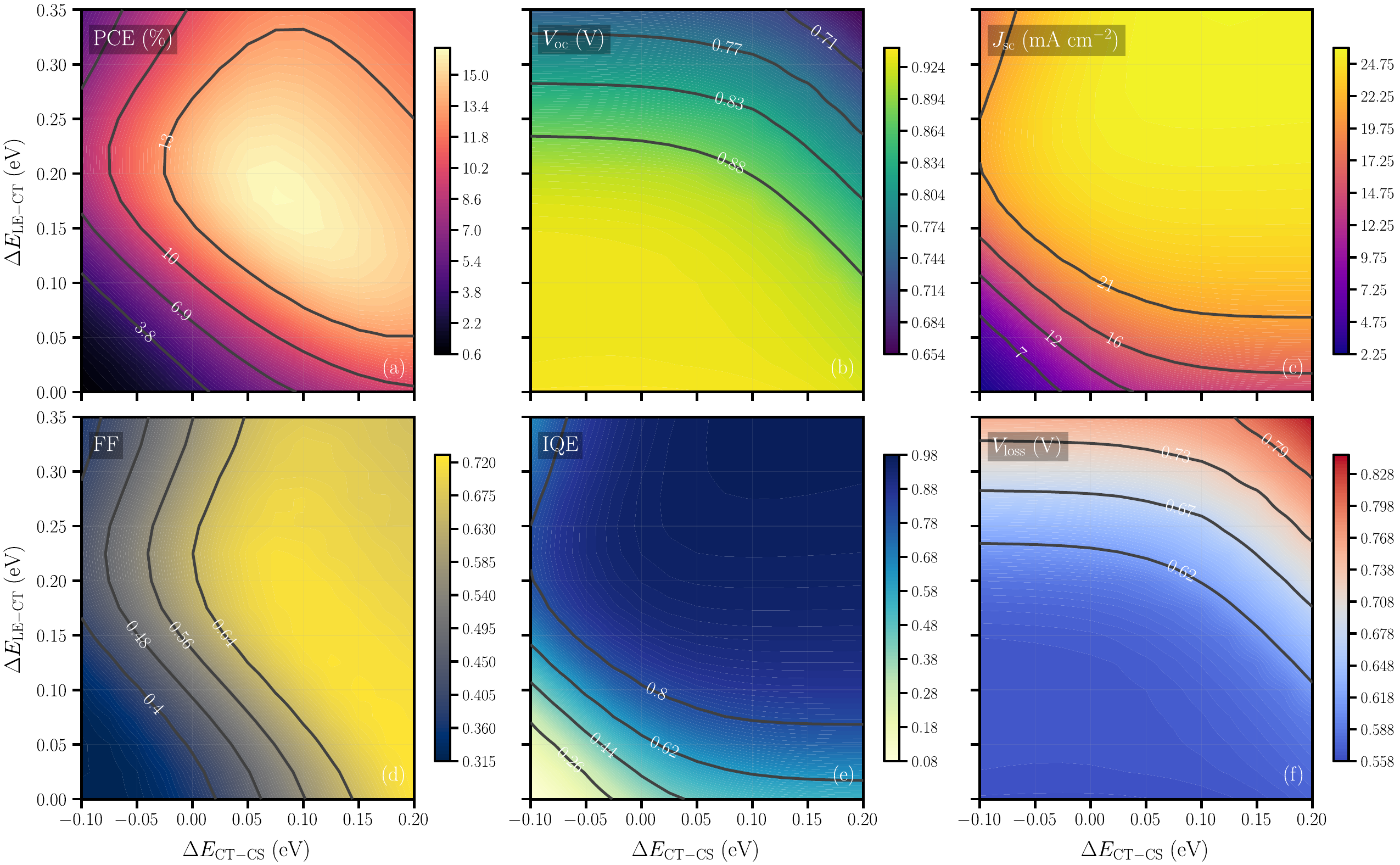}
\caption{\textbf{Two-dimensional EKDD design map for triplet-exciton recycling.} The horizontal axis is $\Delta{}E_{CT-CS}$, and the vertical axis is $\Delta{}E_{LE-CT}$. Panels show (a) PCE, (b) $V_{oc}$, (c) $J_{sc}$, (d) FF, (e) IQE, and (f) $V_{loss}$. Color shading represents the calculated metric values, and contour labels indicate representative numerical levels.}
\label{LE-CT-CS}
\end{figure*}

  Fig.\ref{LE-CT-CS} shows how the initial LE-to-CT step modifies the
  CT--CS trade-off. At small $\Delta{}E_{LE-CT}$, LE-to-CT transfer
  is too slow to compete effectively with LE recombination, producing low IQE
  and $J_{sc}$. Increasing the offset accelerates CT formation and
  improves charge generation until the LE-to-CT step is no longer limiting.
  Beyond this regime, the current benefit saturates, whereas lowering
  $E_{CT}$ and $E_{CS}$ continues to reduce
  $V_{oc}$ and increase $V_{loss}$. Because
  $E_{T}$ is fixed in this scan, increasing the LE--CT offset also
  reduces the CT--T separation and facilitates T to CT3 return. The map
  therefore includes both improved LE-to-CT transfer and improved triplet
  accessibility, but these gains must be balanced against the voltage cost of
  lowering the CT and CS levels.

  The CT--CS dependence remains dominant across the map. A negative,
  endothermic offset suppresses CT3 dissociation even when LE to CT transfer
  is rapid. A large positive offset gives high IQE and $J_{sc}$ but
  sacrifices voltage. The optimum is consequently found at intermediate
  values: PCE reaches 16.24\% at $\Delta{}E_{LE-CT}=0.175$ eV and
  $\Delta{}E_{CT-CS}=0.075$ eV, with $V_{oc}=0.914$ V,
  $J_{sc}=24.87 mA/cm^{-2}$, FF = 0.715, and IQE = 0.941. The
  region within 98\% of the maximum spans approximately
  $\Delta{}E_{LE-CT}=0.15-0.20$ eV and
  $\Delta{}E_{CT-CS}=0.05-0.125$ eV. These ranges identify a
  moderate LE-to-CT driving force that is sufficient for charge transfer
  without unnecessarily lowering the CT and CS energies.

  Taken together, the two maps give a unified energetic design rule. First,
  the CT--CS offset is the primary control parameter and should be slightly
  positive. For the present parameters, the optimum lies near 0.05--0.075 eV:
  CT3 to CS separation is then exothermic enough to generate carriers
  efficiently, but not so strongly downhill that voltage loss dominates.
  Second, the LE--CT offset should be large enough to avoid the
  exciton-transfer bottleneck but remain moderate, approximately 0.15--0.20
  eV here. Third, the CT--T offset should remain small enough for rapid T to
  CT3 recovery; $\Delta{}E_{CT-T}\lesssim{}0.1$ eV, corresponding to
  $K\gtrsim{}10^{-2}$ in the present model, is a useful kinetic
  target.

  These conditions must be optimized jointly because CT is shared by the
  LE--CT--CS and T--CT3--CS pathways. With a negative CT--CS offset, charge
  separation is endothermic and device performance becomes strongly dependent
  on triplet recycling. With an excessively positive offset, direct
  dissociation can maintain the photocurrent and conceal poor recycling, but
  it cannot recover the voltage sacrificed by lowering the CS energy or by
  triplet recombination. High performance therefore requires a moderate
  LE-to-CT driving force, a small positive and weakly exothermic CT--CS
  offset, and a kinetically accessible T to CT3 return pathway.

  \section{Conclusion}

  We developed an exciton-kinetics coupled drift--diffusion (EKDD) model that
  treats LE, CT1, CT3, localized T, and CS populations together with Poisson
  electrostatics and carrier drift--diffusion. The model links microscopic
  excited-state energetics, reversible transfer and recombination, and
  transport to transient extraction, steady-state current--voltage behavior,
  voltage loss, IQE, and PCE. Systematic offset scans show that the three
  energetic controls act at different stages: the LE--CT offset sets the
  onset of interfacial transfer, the CT--CS offset sets the central
  charge-generation/voltage-retention trade-off, and the LE--T offset
  ($\Delta{}E_{ST}$) controls detailed-balance access to triplet
  recycling.

  State-specific parameter scans reveal distinct sensitivities of IQE and
  $V_{loss}$. LE-state properties primarily affect IQE when LE to CT
  transfer is the bottleneck, particularly at small
  $\Delta{}E_{LE-CT}$; once interfacial transfer becomes efficient,
  their influence on IQE is reduced, although they can still modify
  $V_{loss}$ through open-circuit recombination. CT-state properties
  have the strongest direct influence on IQE near zero or negative
  $\Delta{}E_{CT-CS}$, where CT3 to CS dissociation competes with
  CT recombination and triplet formation. At positive CT--CS offsets,
  dissociation becomes dominant and the IQE curves converge, whereas
  $V_{loss}$ remains sensitive because the CS energy and
  open-circuit recombination pathways are changed. T-state properties show a
  different, conditional response: under favorable CT--CS alignment, they
  mainly affect $V_{loss}$ and have a weaker influence on IQE, but
  when CT dissociation is weak or endothermic, T to CT3 recycling becomes
  important and T-state energetics and kinetics can strongly control IQE.
  Thus, LE and CT properties primarily regulate IQE at sequential
  charge-generation bottlenecks, while T properties control the availability
  of triplet recycling; $V_{loss}$ remains broadly sensitive to all
  three states because it depends on the excited-state populations and
  recombination pathways under open-circuit conditions.

  \enlargethispage{2\baselineskip}
  The CT--CS offset is the dominant energetic control. Positive
  $\Delta{}E_{CT-CS}$ makes CT3-to-CS separation exothermic and
  raises $J_{sc}$ and IQE, but lowers the CS energy and increases
  voltage loss. Negative $\Delta{}E_{CT-CS}$ makes separation
  endothermic and preserves the CS energy, but then requires efficient CT3
  dissociation and T-to-CT3 recycling to maintain current. In the D18:FTh-4F
  transient calculation, increasing $\Delta{}E_{ST}$ from 0.15 to
  0.40 eV reduces the recycling ratio by more than four orders of magnitude,
  shortens the delayed extraction tail, lowers the collected pulse fraction
  from 31.1\% to 6.4\%, and decreases steady-state $J_{sc}$ from 28.21
  to 24.33 mA/cm$^{2}$.

  The case studies reproduce terminal triplet-mediated recombination in
  PM6:Y6 and material-dependent recycling in D18:NFA blends. In D18:FTh-4F,
  restoring T-to-CT3 transfer increases $J_{sc}$ from 27.384 to
  28.393 mA/cm$^{2}$ and IQE from 0.9642 to 0.9998, while $V_{oc}$
  increases from 0.7657 to 0.8488 V and $V_{loss}$ decreases from
  0.6443 to 0.5612 V. The two-dimensional maps establish the linked design
  rule: use a moderate LE--CT offset of 0.15--0.20 eV and a small positive,
  weakly exothermic CT--CS offset of 0.05--0.075 eV; simultaneously keep
  CT--T sufficiently small or negative for $K\gtrsim{}10^{-2}$. The
  T--CT3--CS map reaches optimum PCE at $\Delta{}E_{CT-CS}=0.05$ eV
  and $\Delta{}E_{CT-T}=-0.20$ eV, while the LE--CT--CS map reaches
  optimum at $\Delta{}E_{LE-CT}=0.175$ eV and
  $\Delta{}E_{CT-CS}=0.075$ eV. A negative CT--CS offset can
  preserve voltage only when recycling is fast, whereas a strongly positive
  offset can preserve current through direct dissociation while still
  sacrificing voltage. High-performance OPVs therefore require coordinated
  CT--CS thermodynamics, LE to CT transfer, and T to CT3 recycling kinetics.
\section*{ACKNOWLEDGMENTS}
  J.Y. acknowledges funding support from National Natural Science Foundation
  of China (No. 62574175 and No. 62404191), Guangdong Provincial Program (No.
  2023QN10C144), Shenzhen Science and Technology Program (No.
  KQTD20240729102028011 and No. JCYJ20240813113553067), and Guangdong Basic
  Research Center of Excellence for Aggregate Science.
\section*{Author contributions}
  J.Y. conceived the idea. J.D. developed the model. L.L. contributed to the
  exciton decay part of the model. J.D. and  J.Y. wrote the paper. J.Y.
  supervised the work.
\section*{Conflicts of interest}
  The authors declare no competing interest.
\section*{Data availability}
  All data related to the manuscript are available either in the paper or
  through reasonable request from the authors.
\appendix
\section{Impact of LE state}
\begin{figure*}[t]
\centering
  \includegraphics[width=0.93\textwidth]{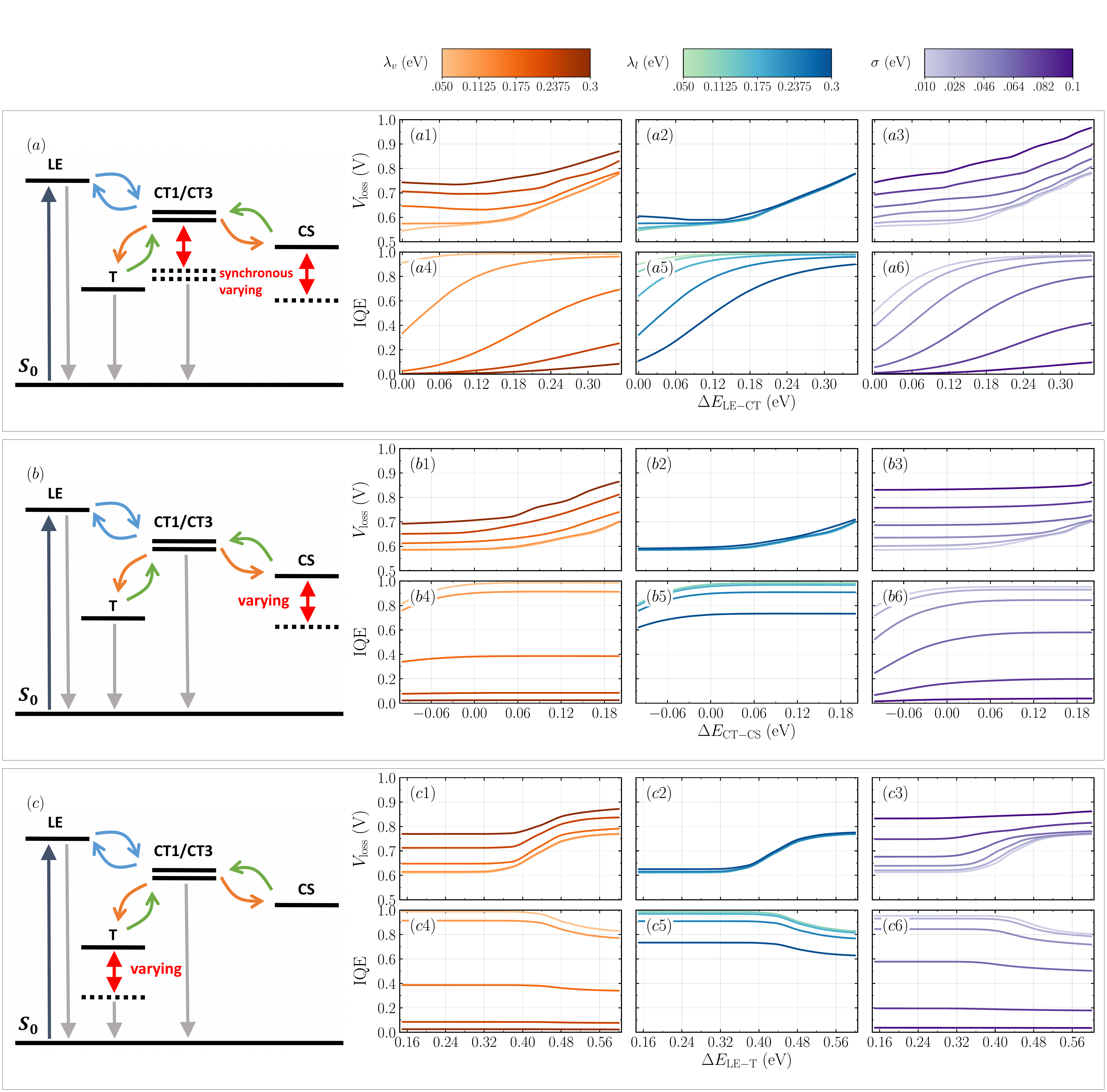}
\caption{Dependence of $V_{loss}$ and IQE on the three energy offsets and LE-state properties. (a1)-(a6): LE--CT offset, (b1)-(b6): CT--CS offset, and (c1)-(c6): LE--T offset. The columns show variations in the T-state high-frequency reorganization energy $\lambda{}_{v}$, low-frequency reorganization energy $\lambda{}_{l}$, and static energetic disorder $\sigma{}$, respectively.}
\label{LE-offset}
\end{figure*}

  The Fig.\ref{LE-offset} summarizes the effect of varying the LE-state
  high-frequency reorganization energy $\lambda{}_{v}$,
  low-frequency reorganization energy $\lambda{}_{l}$, and static
  disorder $\sigma{}$. The parameters were varied over 0.05--0.30 eV
  for the reorganization energies and 0.01--0.10 eV for the disorder.

  The LE-state properties have their strongest effect when LE-to-CT transfer
  is rate-limiting. At small $\Delta{}E_{LE-CT}$, the LE population
  can recombine before reaching the CT state, producing a broad spread in
  IQE. Increasing the LE--CT offset accelerates transfer and generally
  improves IQE, but the simultaneous lowering of the CT and CS energies
  increases $V_{loss}$. The LE-state parameters also affect the
  CT--CS scan because the amount of CT population available for dissociation
  depends on the preceding LE-to-CT conversion step. Once CT dissociation
  becomes efficient, the IQE curves become less separated, whereas
  $V_{loss}$ remains sensitive because the open-circuit population
  continues to depend on excited-state relaxation and recombination.

  The LE--T scan is comparatively less sensitive to LE-state properties under
  the reference CT--CS alignment. This indicates that the LE--T offset does
  not independently determine the charge yield when CT dissociation is
  already efficient. Static disorder produces the broadest low-IQE tails
  because it creates low-energy states that can act as traps or recombination
  centers.

\section{Impact of CT state}
\begin{figure*}[t]
\centering
  \includegraphics[width=0.93\textwidth]{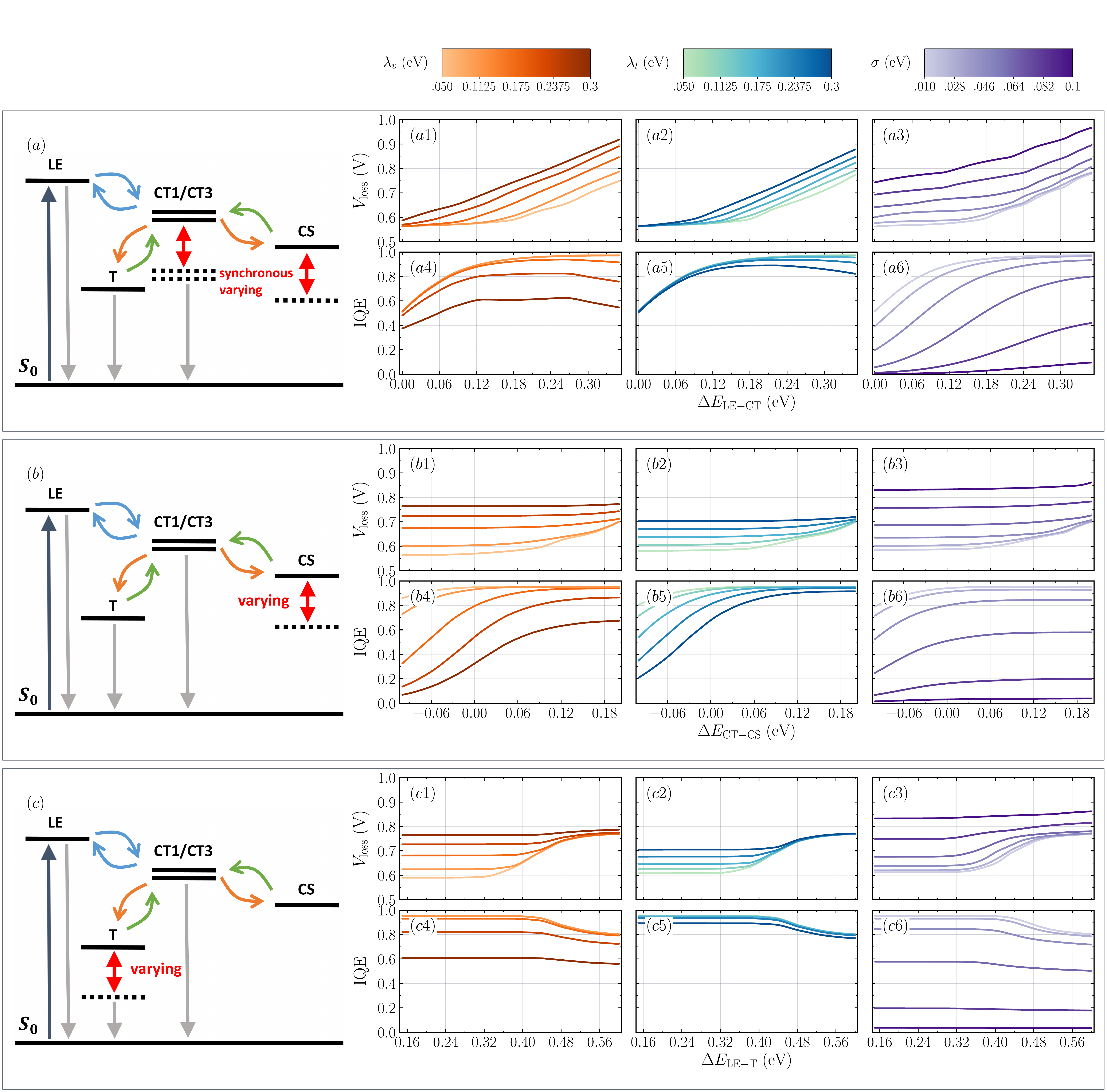}
\caption{Dependence of $V_{loss}$ and IQE on the three energy offsets and CT-state properties. (a1)-(a6): LE--CT offset, (b1)-(b6): CT--CS offset, and (c1)-(c6): LE--T offset. The columns show variations in the T-state high-frequency reorganization energy $\lambda{}_{v}$, low-frequency reorganization energy $\lambda{}_{l}$, and static energetic disorder $\sigma{}$, respectively.}
\label{CT-offset}
\end{figure*}

  The Fig.\ref{CT-offset} shows the corresponding scans for CT-state
  properties. The dominant sensitivity occurs in the CT--CS offset scan,
  particularly near zero and negative $\Delta{}E_{CT-CS}$. In this
  regime, CT dissociation is slow or energetically unfavorable, so changes in
  CT reorganization energies and disorder strongly modify the balance between
  dissociation, recombination, and reverse transfer.

  Positive CT--CS driving force increases the rate of CT dissociation and
  causes the IQE curves to converge. However, because the CS energy is
  lowered when the CT--CS offset is increased in the primary scan, the
  voltage loss increases. This illustrates why the CT--CS offset is more than
  a secondary energetic parameter: it controls both the population flow into
  free carriers and the energetic ceiling for the device voltage.

  The LE--CT scan also exhibits CT-state sensitivity when CT formation and
  subsequent dissociation compete with loss pathways. By contrast, the LE--T
  scan is relatively insensitive to CT properties when the CT--CS offset is
  favorable. In that regime, direct CT dissociation dominates over
  triplet-mediated recycling. The CT-state scans therefore reinforce the
  conclusion that CT--CS alignment is the principal bottleneck governing
  whether changes in excited-state properties affect IQE or mainly appear as
  changes in $V_{loss}$.

\section{Literature-comparison methodology and discussion}
  This method defines the EKDD parameter scan used to generate the shaded
  sensitivity bands in Fig. \ref{EKDD-comp}. The band is a
  model-sensitivity envelope obtained from independently solved device cases;
  it is not a statistical confidence interval or an experimental error bar.

  The reference energies were $E_{LE} = 1.5 eV$, $E_{CT} =
  1.3 eV$, $E_{T} = 1.2 eV$, and $\Delta{}E_{CT-CS} = 0.10 eV$.
  Three offset scans were
  considered. For the LE-CT scan, $E_{LE}$ and $E_{T}$ were
  fixed, $E_{CT} = E_{LE} - \Delta{}E_{LE-CT}$, and
  $E_{CS} = E_{CT} - \Delta{}E_{CT-CS}$, so that the CT-CS
  offset remained fixed. For the CT-CS scan, $E_{LE}$,
  $E_{CT}$, and $E_{T}$ were fixed and $\Delta{}E_{LE-CT}=0.20 eV$.
  For the LE-T scan,
  $E_{CS} = E_{CT} - \Delta{}E_{CT-CS}$, and
  $E_{LE}$, $E_{CT}$ and $E_{CS}$ were fixed and
  $E_{T} = E_{LE} - \Delta{}E_{LE-T}$.

  For each energy alignment, the high-frequency reorganization energy
  $\lambda{}_{v}$ and low-frequency reorganization energy
  $\lambda{}_{l}$ of each of the LE, CT, and T states were
  independently varied over 0.05, 0.1125, 0.175, 0.2375, and 0.30 eV. The
  oscillator strength of each manifold was independently varied over
  $10^{-5}$, $10^{-4}$, $10^{-3}$,
  $10^{-2}$, $10^{-1}$, and 1. Energetic disorder was
  applied as a common Gaussian width to all three states and was varied over
  $\sigma = 0.01$, 0.0575, 0.105, 0.1525, and 0.20 eV. Only
  one parameter and one state were varied in each calculation; all unscanned
  quantities were retained at their reference values. Nine distributed states
  were used for each state. The triplet radiative channel was omitted, such
  that $k_{rec, T} = k_{nr, T}$.

\clearpage
\bibliography{triplet}
\end{document}